\documentstyle[12pt]{article}
\def\lromn#1{\uppercase\expandafter{\romannumeral#1}}

\def\blist{\begin{list}{\setlength{\rightmargin}{\leftmargin}}}
\def\elist{\end{list}}
\makeatletter

\renewcommand{\theequation}%
{\arabic{section}.\arabic{equation}}
\@addtoreset{equation}{section}
\renewcommand{\appendix}{\par
  \setcounter{section}{0}
  \setcounter{subsection}{0}
  \renewcommand{\thesection}{Appendix~\Alph{section}}
  \renewcommand{\theequation}{\Alph{section}.\arabic{equation}}}

\makeatother

\begin{document}

\begin{flushright}
TU/96/506 \\
hep-ph/9608374
\end{flushright}

\vspace{12pt}

\begin{center}
\begin{Large}

\bf{
Quantum System under Periodic Perturbation:
Effect of Environment
}
\end{Large}

\vspace{36pt}

\begin{large}
M. Hotta, I. Joichi, Sh. Matsumoto, and
M. Yoshimura

Department of Physics, Tohoku University\\
Sendai 980-77 Japan\\
\end{large}

\vspace{54pt}

{\bf ABSTRACT}
\end{center}

In many physical situations the behavior of a quantum system is affected
by interaction with a larger environment. We develop, using the method
of influence functional, how to deduce the density matrix of the quantum
system incorporating the effect of environment.
After introducing characterization of the environment by spectral
weight, we first devise schemes to approximate the spectral weight,
and then a perturbation method in field theory models,
in order to approximately describe the environment.
All of these approximate models may be classified as extended Ohmic
models of dissipation whose differences are in the high frequency
part.

The quantum system we deal with in the present work is 
a general class of harmonic oscillators with arbitrary time dependent
frequency.
The late time behavior of the system is well described by an approximation
that employs a localized friction in the dissipative part of the correlation
function appearing in the influence functional.
The density matrix of the quantum system is then 
determined in terms of a single
classical solution obtained with the time dependent frequency.
With this one can compute the entropy, the energy distribution function,
and other physical quantities of the system in a closed form.

Specific application is made to the case of periodically varying 
frequency. This dynamical system has a remarkable property when
the environmental interaction is switched off:
Effect of the parametric resonance gives rise to an exponential growth of
the populated number in higher excitation levels, 
or particle production in field theory models.
The effect of the environment is investigated for this dynamical system
and it is demonstrated that there exists a critical strength of the 
friction for the parametric effect.
In the model of periodically oscillating field coupled to
a system quantum field, it is verified that
the parametric effect occurs in medium, with a somewhat diminished rate,
if the relaxation time scale of the system field towards
thermalization given by the friction term
is larger than the time scale of the coherent parametric amplification.
The effect persists until the back reaction against the periodic oscillation 
stops particle production.
The resulting energy distribution of produced particles described by
a universal function
deviates from the thermal one, having an average energy that
exponentially increases with time.
The dynamical system driven by the parametric oscillator thus
maintains and does not lose its quantum nature even in thermal
bath.
In the present work analytic formulas of 
how physical quantities behave at late times
both in the high and in the low temperature regions are given, along with
results of numerical computation displaying time evolution.



\newpage

\section{Introduction}

\vspace{0.5cm} 
\hspace*{0.5cm} 
Any system of physical interest has interaction with the rest of
a much larger system, and one must clarify effect of the environment
on the system under study.
The problem can be bypassed in some idealized situations in which
the effect of the environment is considered to be too small and 
essentially negligible. 
In the other unique case one deals with the entire universe and in this case
there is no clear separation between the system and the environment.
Except in these special cases one often has a clear idea of how to separate
the small system and the large environment, with some kind of interaction
between them. The important issue one deals with
is then to estimate the effect of environment without knowing too much
of its property, usually difficult to precisely specify.
Even more crucially, in many physical problems
effects of the environment are of central importance.
As examples of such, and only those we come to recognize immediately, we may
list Brownian motion, quantum tunneling and decoherence phenomena.

The first purpose of this paper is to develop a general formalism in 
a special and important subclass of the system-environment interaction:
a quantum system of time dependent harmonic oscillators $q$ 
interacting with the environment variable $Q_{a}$ via bilinear coupling:
\( \:
q\,\sum_{a}\,c_{a}Q_{a}
\: \).
There must exist infinitely many environmental variables distinguished here
by the index $a$ for the environment to behave as a large reservoir.
There already exists a large body of literature that focuses on this
class of models, but we introduce a spectral characterization of the
environment system. In this way we can devise  useful schemes of
approximation for the environment consistently with general principles.

Extension to an infinite set of time dependent harmonic oscillators $q_{k}$
for the system in question 
has a variety of applications in many diverse topics.
This class of models, for instance, includes an interesting case 
of field theory models
in which the system field $\varphi $ interacts with the environment
field $\chi $ via the Yukawa type of interaction Lagrangian density:
\( \:
{\cal L}_{{\rm int}} = -\,\frac{\mu }{2}\,\varphi \,\chi ^{2} \,,
\: \)
with $\mu $ a dimensional coupling constant.
The bilinear composite field $\chi ^{2}$ plays the role of the environmental
variable $Q_{a}$ in this model.
The index $k$ for the system in question is taken here to be the wave vector 
$\vec{k}$ of mode if we are interested in a homogeneous environment.
We propose to study this type of the system-environment interaction
beyond the often-discussed simple model of the quadratic type,
\( \:
\propto \varphi \,\sum_{a}\,c_{a}\,\chi_{a} \,.
\: \)
We have some interesting results for this new class of field theory models
at one loop level.

The second purpose of this paper, and actually the original one that
motivated us towards this problem, is to examine effects of thermal
bath on particle production under periodic perturbation.
There are a few cases that particle production under
periodically oscillating field may become important issues in cosmology.
It was recognized in cosmology that
periodic coherent field oscillation may give a major source of entropy
production in our universe, perhaps the most obvious candidate being 
inflaton oscillation right after inflation.
This specific cosmological problem is 
called the reheating problem \cite{inflation review}.
The essential part of the reheating problem belongs to the class of models 
of time dependent harmonic oscillators set up for the first purpose.
The time dependent frequency in this case is given by a sinusoidal function
oscillating like
\( \:
\cos (m_{\xi }\,t)
\: \).
The inverse period $m_{\xi }$ is the inflaton mass in the reheating problem.
What has been discovered in recent study 
\cite{reheating parametric} --- \cite{kls-2}
of the reheating and related problems is a very rapid and 
efficient mechanism of particle production via the parametric resonance.
The effect drastically changes the old picture \cite{reheat-original} 
of particle production based on the simple inflaton decay.
However, in the usual discussion of the parametric effect 
one ignores the presence of the cosmic medium.
Here are some important, unsolved problems: 
without the environmental interaction 
there is no randomness or entropy among produced particles.
Another application of the parametric resonance in cosmology is decay of very
weakly interacting field such as the Polonyi field, which has a danger of
creating too much entropy when they decay at late times.
In these problems 
it is not immediately clear how the effect of the environment 
affects the rate of particle production, or even the presence of the 
parametric effect itself.
We shall considerably advance our understanding on these problems
in the present work.
In particular, we prove for the first time that the parametric
effect can occur in medium, and give a critical condition for this.

We must stress in the beginning limitation of our approach:
that our formalism is based on the
possibility of the clear cut separation of the system and the environment.
Hence it is difficult for us to discuss the important phase of the reheating
problem: namely when the environment itself rapidly changes by
created particles, ultimately dominating the rest of the environment.
We thus cannot discuss the entire history of reheating.
Nevertheless, we may give a definite picture of how particles are
produced, including their energy distribution at the instant of
their creation prior to interaction among created particles themselves.

On the other hand, 
our formalism is flexible enough to cope with many states of the environment.
For simplicity and as our initial step of investigation, 
we treat the environment in thermal equilibrium characterized by
a finite temperature $T = 1/\beta $. Needless to say, this includes
the quantum ground state as a limit of $T \rightarrow 0$.

The rest of this paper is organized as follows.
In section \lromn 2 we first review briefly the influence functional
method of Feynman and Vernon \cite{feynman-vernon}. We discuss how the correlation function
$\alpha (\tau - s)$ that appears in the influence functional is
identified to the real-time thermal Green's function in the operator
formalism. We then implement this method by
introducing the Lehmann spectral weight for the case of bilinear
coupling between the system and the environment variables.
The spectral weight contains information of both the coupling and
excitation spectrum of the environment, hence we call this the response
weight. The correlation function in the influence functional 
is completely characterized by this response weight.
We organized this section from the point of the inluence functional,
to stress that this method, when supplemented by the spectral
representation, becomes a powerful tool to characterize the system-environment
interaction.

In section \lromn 3
we devise, consistent with some general properties of the response weight,
approximate models of the weight consisting of simple poles of 
relevant analytic functions. This model extends the Ohmic dissipation
model often used in the literature and has the response weight of the
form,
\( \:
r(\omega ) = \frac{\eta }{\pi }\,\omega \,f(\frac{\omega }{\Omega })
\: \).
$\eta $ is the friction constant and the cutoff function 
$f(x) \rightarrow 0$ as $x \rightarrow \infty $.
A general argument is given to support the Ohmic behavior at low
frequencies:
\( \:
r(\omega ) \propto \omega 
\: \)
as $\omega \rightarrow 0$.
The second class of approximations we discuss is perturbation in terms
of the coupling $\mu $ in 
the field theory model given by ${\cal L}_{{\rm int}}$ above.
A full response weight at one loop level is given in this field
theory model.
If the main dissipation of the system field $\varphi $ occurs via
$\chi + \varphi \leftrightarrow \chi $, as so happens typically
at high temperatures, the field theory model may also be regarded as
an extended Ohmic model.
The cutoff function differs much in various models discussed in
this work.
We compute in all of these approximations the friction constant
that becomes important in subsequent sections.

In section \lromn 4 we introduce, as a model of the quantum system,
time dependent harmonic oscillators. 
If one solves this case for generic time dependence in a closed form, 
it should be possible
to deal with much more general class of models for the system.
We explain in detail a crucial assumption for the correlation:
the localized friction approximation for the dissipative part of the
correlation function $\alpha_{I}(\tau ) \sim \eta\, \delta '(\tau )$.
This approximation is excellent for discussion of the late time behavior.
Under this approximation the reduced density matrix of the quantum system
is completely determined in terms of a single classical solution  $u(t)$
of the system, obeying 
\( \:
\stackrel{..}{u} + \omega ^{2}(t)u = 0 \,,
\: \)
with the initial condition,
\( \:
u(0) = 0 \,, \hspace{0.3cm}
\dot{u}(0) = 1 \,.
\: \)
$\omega ^{2}(t)$ here is modified from the original one in a known way.
We explicitly compute the density matrix, the energy distribution function,
and the entropy of the system,
taking as the environment the one in thermal equilibrium.
The energy distribution function, when 
written in terms of a scaled energy divided by the average energy,
is given by a universal function
that has a single spectral shape parameter.
We also check how all this works in the special case of time independent
harmonic oscillators, and compare our result with previous works
\cite{senitzky} --- \cite{b.l. hu et al}.
Some comment on the early time behavior is also made.

In section \lromn 5 we discuss effect of thermal bath on the quantum
system of frequency $\omega $ subjected to periodic oscillation,
\( \:
\xi (t) = \xi _{0}\,\cos (m_{\xi }t)
\: \).
Without the environmental interaction this system exhibits the exponential
growth of population in oscillator excitation levels when the parameters
of the oscillation are in one of the infinitely many bands of
the parameter space $(\omega /m_{\xi } \,, \xi _{0}/m_{\xi })$.
We clarify the condition under which the parametric resonance enhancement
persists despite interaction with the thermal bath:
If the time scale in which a disturbance of the system variable
is relaxed to equilibrium
is larger than the growth time scale of level population, then
the exponential growth occurs with a somewhat diminished rate
compared to that in vacuum.
The condition of the exponential growth also gives a critical
strength of the friction above which the parametric effect does not
occur.
For quantum fields coupled to a periodic field oscillation this
condition limits the momentum range of particles produced by
the parametric resonance.
Furthermore, although the energy distribution of produced particles deviates 
from the thermal one, it has a long tail in the high energy side.
The average energy is shown to exponentially increase with time.
We present analytic formulas in the high and in the low temperature regions,
as well as results of detailed numerical computation including time
evolution of physical quantities.

Section \lromn 6 gives a summary of the present work.
Some technical detail related to numerical computation is relegated to
Appendix.

\vspace{1cm}
\section{
General framework: influence functional 
and spectral representation}

\vspace{0.5cm} 
\hspace*{0.5cm} 
The Lagrangian of our problem consists of three parts: using the 
system variable $q$ and the environment variable $Q$, it is
\begin{equation}
L = L_{q}[q] + L_{{\rm int}}[q\,, Q] + L_{Q}[Q] \,.
\end{equation}
We take for the system-environment interaction the bilinear term:
\begin{eqnarray}
L_{{\rm int}} = -\, q\,\sum_{a}\,c_{a}Q_{a} \,.
\end{eqnarray}
In later discussions we mainly focus on the time dependent
harmonic oscillator for the system dynamics,
but for the time being, namely in the present and the next sections,
we may choose any $L_{q}$ for the system dynamics.
Since we do not specify the nature of the environment variable 
except that $Q_{a}$'s are independent,
the bilinear coupling $L_{{\rm int}}$ above may be regarded as a general one.
For instance, the environment variable $Q_{a}$ may be a composite operator 
made of more fundamental fields,
as discussed in the field theory model in the next section.
In any case it is important to have infinitely many indices $a$ coupled to
the single system variable $q$, because the environment must behave as a
large system compared to the $q-$system.

It is very useful to introduce the influence functional, following
Feynman and Vernon \cite{feynman-vernon}. 
The basic idea is that one is interested in the
behavior of the $q-$system alone
and traces out the environment variable altogether.
We define here the influence functional by convoluting with the initial
state of the environment. To do so we assume that initially we may take
an environment state uncorrelated with the system.
In the path integral framework the influence
functional is thus obtained after integrating out the environment variables:
\begin{eqnarray}
\hspace*{-0.5cm}
{\cal F}[\,q(\tau )\,, q'(\tau )\,] &\equiv& 
\int\,{\cal D}Q(\tau )\,\int\,{\cal D}Q'(\tau )\,\int\,dQ_{i}\,\int\,dQ'_{i}
\int\,dQ_{f}\,\int\,dQ'_{f}\,\delta (Q_{f} - Q'_{f})\,
\nonumber \\
\hspace*{-0.5cm} && 
K \left(\,q(\tau )\,,Q(\tau ) \,\right)\,
K^{*} \left( \,q'(\tau )\,,Q'(\tau ) \,\right)\,\rho_{i} (Q_{i}\,, Q'_{i}) \,,
\\
\hspace*{-0.5cm}
K \left( \,q(\tau )\,,Q(\tau ) \,\right) &=& 
\exp \left( \,iS_{0}[Q] + iS_{{\rm int}}[q \,, Q]\,\right) \,, \\
\hspace*{-0.5cm}
S_{0}[Q] + S_{{\rm int}}[q \,, Q] &=& \int_{0}^{t}\,d\tau \,
\left( \,L_{Q}[Q] + L _{{\rm int}}[q\,, Q]\,\right) \,.
\end{eqnarray}
It is a functional of the entire path of the system $q(\tau )$ and
its conjugate $q'(\tau )$.
\( \:
\rho_{i} (Q_{i}\,, Q'_{i})
\: \)
is the initial density matrix of the environment, which can be any
mixture of pure quantum states.
What deserves to be stressed is that one does not observe the final
state of the environment, hence integration with respect to the final
values of $Q_{f} = Q'_{f}$ is performed here.

Once the influence functional is known, one may compute the transition
amplitude and any physical quantities of the $q-$system
by convoluting dynamics of
the system under study. For instance, the transition amplitude is given by
\begin{eqnarray}
&&
\int\,{\cal D}q(\tau )\,\int\,{\cal D}q'(\tau )\,
\int\,dq_{i}\,\int\,dq'_{i}\,\int\,dq_{f}\,\int\,dq'_{f} \nonumber \\
&& \hspace*{1cm} 
\psi ^{*}(q_{f})\,\psi ^{*}(q'_{i})\,{\cal F}[\,q(\tau )\,, q'(\tau )\,]\,
e^{iS[q] - iS[q']}\,
\psi (q_{i})\,\psi (q'_{f}) \,,
\end{eqnarray}
where $\psi $'s are wave functions of the initial and
the final $q-$states and $S[q]$ is the action of the $q-$system.

The form of the influence functional is dictated by general principles
such as probability conservation and causality. Feynman and Vernon 
found a closed quadratic form consistent with these,
\begin{eqnarray}
{\cal F}[\,q(\tau )\,, q'(\tau )\,] &=& \nonumber \\
&& \hspace*{-3cm}
\exp \left[\,-\,\int_{0}^{t }\,d\tau \,\int_{0}^{\tau }\,ds\,
\left( \,\xi (\tau )\alpha_{R}(\tau - s)\xi (s) + i\,\xi (\tau )
\alpha _{I}(\tau - s)X(s)\,\right)\,\right] \,, 
\label{influence-f def} \\
&& \hspace*{-2cm}
{\rm with} \;
\xi (\tau ) = q(\tau ) - q'(\tau ) \,, \hspace{0.5cm} 
X(\tau ) = q(\tau ) + q'(\tau ) \,.
\end{eqnarray}
Thus two real functions $\alpha _{i}(\tau )$ are all we need to
characterize the system-environment interaction.
These are defined here in the range of 
\( \:
\tau \geq 0 \,.
\: \)
The fact that $\alpha_{i} $ depends on the difference of time
variables, $\tau - s$, is due to the assumed stationarity of the environment.
Without this stationarity these may depend on $\tau \,, s$ individually:
$\alpha _{i}(\tau \,, s)$.

It would be useful to recall that if one takes as the environment
an infinite collection of independent
harmonic oscillators in thermal equilibrium of temperature
$T = 1/\beta $, then
\begin{eqnarray}
\alpha _{R}(\tau) &=& \sum_{a}\,c_{a}^{2}\,\alpha _{R}^{(a)}(\tau ) \,, 
\hspace{0.5cm} 
\alpha _{I}(\tau) = \sum_{a}\,c_{a}^{2}\,\alpha _{I}^{(a)}(\tau) \,, \\
\alpha _{R}^{(a)}(\tau ) &=& \coth (\frac{\beta\omega_{a}}{2}) \,
\frac{\cos (\omega_{a}\,\tau )}{2\omega_{a} }\,,
\\
\alpha _{I}^{(a)}(\tau ) &=& 
-\,\frac{\sin (\omega_{a}\,\tau )}{2\omega_{a} }\,{\rm sign}\,(\tau) \,, 
\end{eqnarray}
with sign$\,(\tau )$ the signature function,
\( \:
{\rm sign}\,(\tau ) = 1 \; (\tau > 0) \,, \; = -1 \; (\tau < 0) \,.
\: \)
We extended the $\tau $ range to $-\infty < \tau < \infty $ such that
the function in $\tau < 0$ is consistent with eq.(\ref{influence kernel})
later discussed;
\( \:
\alpha _{i}^{(a)}(-\,\tau ) = \alpha _{i}^{(a)}(\tau )   \,.
\: \)
If one Fourier-transforms 
\begin{equation}
\alpha^{(a)} (\tau ) = \alpha _{R}^{(a)}(\tau ) + i\alpha _{I}^{(a)}
(\tau ) \,, 
\end{equation}
with
\begin{equation}
\alpha^{(a)} (\omega ) \equiv \int_{-\infty }^{\infty }\,d\tau \,\alpha^{(a)}
(\tau )\,e^{i\omega \tau } \,, 
\end{equation}
then one finds the familiar form,
\begin{eqnarray}
\alpha^{(a)} (\omega ) = i\,\left( \frac{1}{\omega^{2} -
\omega_{a}^{2} + i\epsilon } - \frac{2\pi i}{e^{\beta \omega _{a}} - 1}
\,\delta (\omega^{2} - \omega _{a}^{2}) \right) \,.
\end{eqnarray}

Emergence of the thermal propagator is by no means an accident. 
By comparing the path integral formula and the operator
formula, one can indeed derive the equivalence of $i\alpha^{(a)} (\tau )$
to the real-time thermal Green's function;
\begin{equation}
\alpha^{(a)} (\tau ) = {\rm tr}\;\left( \,\rho _{\beta }\,
T[\,\tilde{Q}_{a}(\tau )\,\tilde{Q}_{a}(0)\,]\,\right) \,.
\label{influence kernel} 
\end{equation}
$\rho _{\beta }$ is the thermal density matrix 
\( \:
e^{-\,\beta H_{Q}}/(\,{\rm tr}\; e^{-\,\beta H_{Q}}\,) \,
\: \)
of the environment.
In order to avoid confusion, we explicitly denoted the environment 
operators by the tilde notation.
The key point for this identity is the form of the bilinear coupling,
\( \:
L_{{\rm int}} = -\,q\,\sum_{a}\,c_{a}Q_{a} \,.
\: \)
Taking as the initial environment an arbitrary quantum state,
not necessarily limited to the thermal state, gives
the influence functional of the form,
\begin{eqnarray}
{\cal F}[\,q(\tau )\,, q'(\tau )\,] &=& \int\,{\cal D}Q(\tau )\,
\int\,{\cal D}Q'(\tau )\,
\int\,dQ_{i}\,\int\,dQ_{i}\,'\,\rho _{i}(Q_{i}\,, Q_{i}\,') 
\nonumber \\
&&
\int\,dQ_{f}\,\int\,dQ_{f}\,'
\,\delta (Q_{f} - Q_{f}\,')\, 
\exp\, \left[ \,i\left(\, S_{0}[Q] - S_{0}[Q'] \,\right) \right.
\nonumber \\
&&
\left. -\, \frac{i}{2}\,\int_{0}^{t}\,d\tau \,
\sum_{a}\,c_{a}\left( \,\xi(\tau )\, (Q + Q')_{a}(\tau ) + 
X(\tau )\,(Q - Q')_{a}(\tau )\,\right)\,\right] \,, \nonumber \\
\end{eqnarray}
which is to be equated to
\begin{eqnarray*}
\exp [\,- \,\int_{0}^{t}\,d\tau \,\int_{0}^{\tau }\,ds\,
\xi (\tau )\left( \,\alpha _{R}(\tau \,, s)\xi (s) + i\alpha _{I}
(\tau \,, s)X(s) \,\right)\,] \,. 
\end{eqnarray*}
Here
\begin{equation}
S_{0}[Q] = \int_{0}^{t}\,d\tau \,L_{Q}(Q)
\end{equation}
is the action of the environment.
Functional derivatives with respect to 
$\xi (\tau )$ and $X(\tau )$
yield various combination of expectation values of the operators,
\( \:
\tilde{Q}(\tau ) \,\pm\, \tilde{Q}'(\tau ) \,.
\: \)
A suitable combination of these leads to the required identity 
(\ref{influence kernel}).

It should be evident that the fundamental relation
(\ref{influence kernel}) holds for any initial $Q-$state if
one takes the relevant density matrix $\rho_{i} $.
Equivalence of the path integral formula to the corresponding operator
formula is used to extract useful characterization of the environment, 
as will be explained.

We now introduce the Lehmann spectral representation. Without explicitly
solving interaction with the environment,
this spectral knowledge is adequate to predict the behavior of the system
under study.
Respecting some general properties of the spectral function given below,
one may also devise useful approximate schemes for the spectral density. 
This approach is thus very useful to characterize the environment
without knowing too much of it.

Assuming the thermal density matrix for the initial $Q-$system, one 
obtains, by using
\( \:
\tilde{Q}_{a}(\tau ) = e^{iH_{Q}\tau }\,\tilde{Q}_{a}(0)\,e^{-\,iH_{Q}\tau }
\: \) 
and inserting a complete set of eigenstates $|\,n\,\rangle $
of the Hamiltonian $H_{Q}$, the thermal real-time Green's function:
\begin{eqnarray}
&&
G(\tau ) = \sum_{a}\,c_{a}^{2}\,G_{a}(\tau ) \,, 
\\
&&
i\,G_{a}(\tau ) = {\rm tr}\,\left( \,
T[\,\tilde{Q}_{a}(\tau )\,\tilde{Q}_{a}(0)\,]
\,\rho _{\beta }\,\right) = \frac{1}{\sum_{l}\,e^{-\beta E_{l}}} 
\nonumber \\
&&
\cdot
\sum_{n \,, m}\,|\,\langle n|\tilde{Q}_{a}(0)|m \rangle\,|^{2}\,
e^{-\beta E_{n}}
\,\left( \,e^{-i(E_{m} - E_{n})\,\tau }\,\theta (\tau )
+ e^{i(E_{m} - E_{n})\,\tau }\,\theta ( - \tau )\,\right) \,.
\end{eqnarray}
The Fourier transform defined by
\begin{equation}
G_{a}(\omega ) \equiv \frac{1}{2\pi }\,\int_{-\infty }^{\infty }\,d\tau \,
G_{a}(\tau )e^{i\omega \tau } 
\end{equation}
yields with $\omega _{mn} = E_{m} - E_{n}$
\begin{eqnarray}
\hspace*{-1cm}
G_{a}(\omega ) &=& \frac{1}{2\pi }\,
\sum_{n \,, m}\,|\,\langle n|\tilde{Q}_{a}(0)|m \rangle\,|^{2}
\nonumber \\
&&
\cdot e^{-\beta E_{n}}
\,\left( \,\frac{1}{\omega + i\epsilon - \omega _{mn}} - 
\frac{1}{\omega - i\epsilon + \omega _{mn}} \,\right)\,
\frac{1}{\sum_{l}\,e^{-\beta E_{l}}} 
\\
\hspace*{-1cm} &=& 
\frac{1}{2\pi }\,\int_{-\infty }^{\infty }\,d\omega '\,
\left( \,\frac{1}{1 - e^{-\beta \omega }}\,\frac{r_{a} (\omega ')}
{\omega - \omega ' + i\epsilon } + \frac{1}{1 - e^{\beta \omega }}\,
\frac{r_{a} (\omega ')}{\omega - \omega ' - i\epsilon }\,\right) \,, 
\\
\hspace*{-1cm}
r_{a} (\omega ) &\equiv& (1-e^{-\beta \omega })\,
\sum_{n \,, m}\,|\,\langle n|\tilde{Q}_{a}(0)|m \rangle\,|^{2}\,
e^{-\beta E_{n}}
\,\delta (\omega - \omega _{mn})\,\frac{1}{\sum_{l}\,e^{-\beta E_{l}}} \,.
\end{eqnarray}
The quantity
\begin{equation}
r(\omega ) = \sum_{a}\,c_{a}^{2}\,r_{a}(\omega )
\label{response weight def} 
\end{equation}
we call the response weight, in this paper.

We shall discuss general properties of the response weight $r(\omega )$
shortly. It would be instructive to recall beforehand for a single harmonic 
oscillator for $Q$,
\begin{equation}
r_{a}(\omega ) = \frac{1}{2\omega _{a}}\,
\left(\,\delta (\omega - \omega _{a}) - \delta (\omega + \omega _{a})\,\right)
\,.
\end{equation}
This can be computed by using
\( \:
\tilde{Q}_{a}(0) = \frac{1}{\sqrt{2\omega _{a}}}\,(a_{k} + a_{k}^{\dag })
\: \)
in terms of the creation and the annihilation operators of harmonic
oscillator and summing over the energy levels,
\( \:
E_{n} = (n + \frac{1}{2})\,\omega _{a} \,.
\: \)
Despite the initial appearance of the temperature factor $\beta $ it
disappears in the final form of the response weight, as expected from
previous results.
For a collection of harmonic oscillators in thermal equilibrium,
the response weight is thus
\begin{equation}
r(\omega ) \equiv \sum_{a}\,c_{a}^{2}\,
\delta (\omega ^{2} - \omega _{a}^{2})\,{\rm sign}\,(\omega ) =
\sum_{a}\,\frac{c_{a}^{2}}{2\omega _{a}}\,\left(\,
 \delta (\omega - \omega _{a})
- \delta (\omega + \omega _{a}) \,\right) \,.
\label{ho response weight} 
\end{equation}
The response weight in this form was introduced in \cite{caldeira-leggett 83},
but we stress that here it is defined in the most general way,
using the spectral representation for arbitrary environment.

The response weight $r(\omega )$ thus defined characterizes
the response of elementary excitation in thermal environment
that may couple to the system variable.
Thus only this quantity needs to be specified to determine the behavior
of the $q-$system. By definition it obeys a positivity constraint,
\begin{equation}
r(\omega )\,{\rm sign}\,(\omega ) \geq 0 \,.
\end{equation}
Furthermore we often assume as an acceptable system of environment
that the high frequency limit of the response weight is bounded such that
\begin{equation}
r(\omega ) \:\rightarrow  \: 
O[\,\omega ^{-\,\epsilon }\,] \hspace{0.3cm} {\rm as} \; 
\omega \:\rightarrow  \: \infty  \,,
\label{asymptotic response} 
\end{equation}
with $\epsilon > 0$.
If this condition is not satisfied, one must use instead of $r(\omega )$
a better behaved
\( \:
r(\omega )/\omega ^{n}
\: \)
with appropriate $n\geq 1$ in subsequent dispersion integrals, and
must introduce as a subtraction term a polynomial in $\omega $ of order
$\leq n - 1$.
It is important to note that the response weight is an imaginary part of some
analytic functions. With the 
usual type of retarded and advanced Green's functions,
\begin{eqnarray}
&&
G(\omega ) = \frac{1}{1 - e^{-\beta \omega }}\,G^{R}(\omega ) +
\frac{1}{1 - e^{\beta \omega }}\,G^{A}(\omega ) 
\label{fourier green} 
 \,, \\
&&
G^{R\,, A}(\omega ) = \int_{-\infty }^{\infty }\,\frac{d\omega '}{2\pi }\,
\frac{r(\omega ')}{\omega - \omega ' \pm i\epsilon } = 
\int_{0}^{\infty }\,\frac{d\omega '}{\pi }\,
\frac{\omega '\,r(\omega ')}
{\omega ^{2} - \omega '\,^{2} \pm i\epsilon \omega } \,,
\end{eqnarray}
one has
\begin{eqnarray}
r(\omega ) = 2\,\Im \,G^{A}(\omega ) = -\,2\,\Im \,G^{R}(\omega ) \,.
\end{eqnarray}
The retarded (advanced) $G^{R}(\,G^{A}\,)$ is analytic 
in the upper (lower) half $\omega-$plane.

A sum rule on the spectral function is usually taken granted.
This holds if $Q_{a}$ is a canonical variable, and
takes the form of the normalization integral,
\begin{equation}
\int_{-\infty }^{\infty }\,d\omega \,\omega \,r_{a}(\omega ) \,.
\end{equation}
We do not consider this constraint any further, because our response weight
(\ref{response weight def}) 
contains the unknown coupling $c_{a}^{2}$ and an unspecified number of
dynamical variables. 

Although the real-time formalism is a natural consequence of the
influence functional method, it is sometimes more convenient to
use the imaginary-time formalism.
If $r(\omega )$ obeys the convergence condition, 
eq.(\ref{asymptotic response}), then there exists a unique
analytic function $\overline{G}(z)$ of a complex variable $z$,
\begin{equation}
\overline{G}(z) = \int_{-\infty }^{\infty }\,\frac{d\omega '}{2\pi }\,
\frac{r(\omega ')}{z - \omega '} \,, 
\end{equation}
that gives the real-time thermal Green's function $G^{R\,, A}(\omega )$ 
for real $\omega $ as limiting functions: 
\begin{equation}
G^{R}(\omega ) = \overline{G}(\omega + i\epsilon ) \,, \hspace{0.5cm} 
G^{A}(\omega ) = \overline{G}(\omega - i\epsilon ) \,.
\end{equation}
The imaginary-time thermal Green's function $\tilde{G}$ defined at discrete
$\omega _{n}$ is simply
\begin{equation}
\tilde{G}(\omega_{n} ) = \overline{G}(\omega _{n}) \,,
\end{equation}
with
\( \:
\omega _{n} \equiv i\,2\pi \,n/\beta \,,
\hspace{0.3cm}
n = \pm 1\,, \pm 2 \,, \cdots \,.
\: \)
This is a well known result \cite{fetter-walecka}.
We note that with the uniqueness of the analytic function, 
the analytically continued imaginary-time Green's function
gives a convenient means of computing the response weight as a discontinuity:
\( \:
r(\omega ) = \left( \, \overline{G}(\omega - i\epsilon )
- \overline{G}(\omega + i\epsilon )\,\right)/i
\: \).
We shall use this fact to compute the response weight in field theory models.

With this machinery the correlation function that appears in
the Feynman-Vernon formula is given by
\begin{eqnarray}
\alpha _{I}(\tau ) &=& -\,\frac{i }{2}\,\int_{-\infty }^{\infty }\,
d\omega \,r(\omega )\,e^{-i\omega \tau }\,{\rm sign}\,(\tau ) 
\nonumber \\
&=& -\,\int_{0}^{\infty }\,d\omega \,r(\omega )\,
\sin (\omega \tau )\,{\rm sign}\,(\tau) \,, \label{imaginary correlation} 
\\
\alpha _{R}(\tau ) &=& 
\frac{1}{2}\,\int_{-\infty }^{\infty }\,d\omega \,
\coth \frac{\beta \omega }{2}\,r(\omega )\,e^{-i\omega \tau } 
\nonumber \\
&=& \int_{0}^{\infty }\,d\omega \,r(\omega )\,\coth 
\frac{\beta \omega }{2}\,\cos (\omega \tau ) \,.
\end{eqnarray}
Their Fourier transforms give the familiar dispersion relation at
finite temperatures,
\begin{eqnarray}
\alpha _{R}(\omega ) &=& \pi \,\coth \frac{\beta \omega }{2}\,r(\omega )
\,, 
\\
\alpha _{I}(\omega ) &=& \frac{2}{\pi }\,{\cal P}\,
\int_{0}^{\infty }\,d\omega '\,
\frac{\omega '\,\tanh \frac{\beta \omega '}{2}\,\alpha _{R}(\omega ')}
{\omega ^{2} - \omega'\, ^{2}} \,.
\end{eqnarray}

In the next section we shall discuss two useful approximate schemes
for the response weight $r(\omega )$, one based on simple
meromorphic forms of this function and the other based on perturbation
in field theory models.

\vspace{1cm}
\section{
Models for environment
}

\vspace{0.5cm} 
\hspace*{0.5cm} 
It is usually difficult or even impossible to know details of the
environment, yet one must be able to extract essential features of
the reservoir in order to predict the behavior of the system under study.
Based on the response weight introduced in connection with the 
influence functional in the preceding section, 
we discuss two approximate schemes to evaluate this quantity. 
These two methods are complimentary. In one method we
follow the spirit of the S-matrix theory: Instead of computing the response
weight from fundamental microphysics 
we approximate relevant analytic functions in terms of
simple pole terms whose parameters may be used to characterize the
environment system. In the other method we solve the fundamental
problem by perturbation theory. What one can compute in this case
is necessarily limited, for instance only the lowest order computation
is readily tractable.
Nevertheless one may gain some insight of how to relate macro-quantity
to microphysics parameters.
There is also a freedom of how to model the environment.
We introduce non-linear models of the system-environment interaction
in field theories. This type of models seem to have a variety of areas
of application, some of which we hope to cover in subsequent works.

\vspace{0.5cm} 
{\bf \lromn 3 A. \hspace{0.3cm}
Meromorphic approximation for the response weight
}

\vspace{0.5cm} 
The critical observation to devise useful scheme of approximation
is that the response weight $r(\omega )$ defined for $\omega $ real
is a limit of some analytic function, 
\begin{equation}
r(\omega ) = \lim_{z \rightarrow \omega ^{+}}\,\Im \,f(z) \,, 
\end{equation}
where $f(z) = -\,2\,G^{R}(z) $ is analytic in the upper half complex $z-$plane.
The simplest approximation for $G^{R}(z)$ is to replace its complicated
analytical
structure by a set of simple poles consistent with general principles.
Moreover, to extract the late time behavior of the correlation function
$\alpha (\tau )$ it is sufficient to determine the low frequency
limit of the response weight $r(\omega )$, as will be explained.
The minimum set of poles thus might be a good approximation.
Even the more complicated case of singularities such as
the branch cut  may be described by a continuous distribution
of poles.

It turns out that the oddness and the positivity on the positive real axis 
of $\omega $ constrains much of pole parameters of 
the response weight $r(\omega )$. 
Poles thus appear in a pair at $z = w _{1}$ and $w_{2}$ where 
\( \:
w _{2} = -\,w _{1}^{*} \,,
\: \)
with residues $c_{i}$ related by
\( \:
c_{2} = -\,c_{1}^{*} \,.
\: \)
Using the minimum set of poles, one has
\begin{eqnarray}
r^{(4)}(\omega ) &=& ic\left(\,
\frac{1}{\omega - w} + \frac{1}{\omega + w} - 
\frac{1}{\omega - w^{*}} - \frac{1}{\omega + w^{*}}
\, \right) \nonumber 
\\
&=&
\frac{4c\Omega \gamma\, \omega }{(\omega^{2} - \Omega ^{2}
+ \frac{\gamma ^{2}}{4})^{2} + \Omega ^{2}\gamma ^{2}} \,, 
\end{eqnarray}
with $\Re w \neq 0$, and for
$\Im w < 0 \,, $
\( \:
\Re c > 0 
\: \) and
\( \:
 \Im c = 0 \,.
\: \)
There are three real parameters, 
\( \:
\Re c \,, \; \Re w \equiv \Omega \,, \; \Im w \equiv -\,\gamma/2 \,.
\: \)
We call this approximation a quartet model. 

The exception occurs for $\Re w = 0$;
\begin{eqnarray}
r^{(2)}(\omega ) &=& 
\left( \,\frac{1}{\omega - i \gamma/2}
+ \frac{1}{\omega + i \gamma/2}\,\right)\,a
\nonumber \\
&=&
\frac{2a\omega }{\omega ^{2} + \gamma ^{2}/4}
\,.
\end{eqnarray}
We call this case a doublet model.
In the literature it is called the Drudes model,
and has been repeatedly discussed \cite{lindenberg-west}.
We shall not treat this model in any detail.

From this response weight the correlation function $\alpha _{i}$
that appears in the Feynman-Vernon formula becomes
\begin{eqnarray}
&& 
\alpha _{R}^{(i)}(\omega ) = \pi \,\coth \frac{\beta \omega }{2}\,
r^{(i)}(\omega ) \,, \\
&&
\alpha _{I}^{(4)}(\tau ) = -\,2\pi \,c\,e^{-\,\gamma \,|\tau |/2}
\,\sin (\Omega \,|\tau |) \,, \hspace{0.5cm} 
\alpha _{I}^{(2)}(\tau ) = -\,\pi \,a\,e^{-\,\gamma \,|\tau |/2}
\,, 
\\
&& 
\alpha ^{(4)}_{R}(\tau ) = 4 c\Omega \gamma \,
\int_{0}^{\infty }\,d\omega \,
\coth  \frac{\beta \omega}{2}\,\frac{\omega }{(\omega ^{2} - \Omega ^{2}
+ \frac{\gamma ^{2}}{4})^{2}
+ \Omega ^{2}\gamma ^{2}}\,\cos (\omega \tau )\,,  \\
&& 
\alpha ^{(2)}_{R}(\tau ) = 2 a\,
\int_{0}^{\infty }\,d\omega \,
\coth  \frac{\beta \omega}{2}\,\frac{\omega }{\omega ^{2} 
+ \gamma^{2}/4}\,\cos (\omega \tau ) \,.
\end{eqnarray}

To gain more physical insight, it is useful to explore the late time 
behavior at
\( \:
t \gg 1/\Omega \hspace{0.3cm}( {\rm with}\; \Omega \gg  \gamma )
\: \)
for the quartet model.
As $\tau \rightarrow \infty $,
$\alpha _{I}^{(4)}(\tau )$ behaves as $\eta \,\delta '(\tau )$
with
\begin{equation}
\eta = -\,2\,\int_{0}^{\infty }\,d\tau \,\tau \,\alpha _{I}^{(4)}(\tau )
\,.
\end{equation}
The leading asymptotic behavior of the correlation function 
$\alpha _{I}(\tau )$ in the quartet model is thus given by 
\begin{equation}
\alpha_{I} ^{(4)}(\tau ) \sim 
\eta\, \delta '(\tau ) \,, \hspace{0.5cm} 
\eta = \frac{4\pi c\gamma \Omega }{(\,\Omega ^{2} + \gamma ^{2}/4\,)^{2}} 
\,. 
\end{equation}
Putting these forms of $\alpha _{I}(\tau )$ 
into the Feynman-Vernon formula (\ref{influence-f def}), 
one gets the friction term of the form,
\begin{equation}
\eta \,q(\tau )\frac{d}{d\tau }q(\tau ) \,,
\end{equation}
ignoring a surface term.
We shall treat the surface term rigorously in later applications.

The quartet model actually contains the frequency shift term 
\( \:
-\,\mu ^{2}q^{2}(\tau )
\: \)
as
a subleading asymptotic term, which is difficult to extract unambiguously.
The frequency shift turns out to correspond to the mass renormalization
in field theory, and is not an observable quantity.
On the other hand, the localized friction term above is very important,
and with the localized
approximation for $\alpha _{I}(\tau )$ above the real part of
correlation is given by
\begin{eqnarray}
&&
\alpha _{R}^{(4)}(\tau ) = \int_{0}^{\infty }\,
d\omega \,\coth (\frac{\beta \omega }{2})\,r^{(4)}(\omega ) 
\,\cos (\omega \tau )
 \,, \\
&&
r^{(4)}(\omega ) \sim  \omega \,r'\,^{(4)}(0) 
= \frac{\eta }{\pi }\,\omega  \,.
\end{eqnarray}
This approximation for the response weight $r(\omega )$ 
is often adequate in determining the behavior at asymptotic late times.
But the frequency integral on $\omega $ is not convergent unless
one uses the full form of $r(\omega )$ such as the one 
given in the quartet model.
We shall discuss the high frequency cutoff later.

The form of the response weight, $r(\omega ) \propto \omega $, is often
called the Ohmic model of dissipation in the literature. 
In this and the next section we are giving our microscopic foundation of
this form of the behavior modified at some high frequency cutoff
$\Omega $.
Although the Ohmic model does not come out of fundamental principles,
its low frequency behavior appears in a variety of interesting models.
To show this we first consider a class of the response weight that 
comes out of the harmonic oscillator bath;
\begin{equation}
r(\omega ) = \frac{c^{2}(\omega )}{2\omega }\,D(\omega ) \,.
\end{equation}
We replaced the discrete sum in the previous formula 
eq.(\ref{ho response weight}) by
a continuous integral introducing the density of states:
$D(\omega )$.
In many systems of physical interest both in condensed matter and
particle physics the density of states behaves as 
\( \:
D(\omega ) \propto \omega ^{2}
\: \)
at $\omega \sim 0$. With $c(\omega ) \rightarrow $ a constant as
$\omega \rightarrow 0$, this means the Ohmic behavior,
$r(\omega ) \propto \omega $.
In analytic approach to the response weight one may consider
approximate models in which poles are continuously distributed;
\begin{equation}
r(\omega ) = \frac{\eta }{\pi }\,\omega \,\int_{\Omega _{0}}^{\infty }\,
d\Omega \,
f(\frac{\omega }{\Omega }) \,.
\end{equation}
We assume the convergence of the $\Omega $ integral.
The cutoff function $f(\frac{\omega }{\Omega })$ can be taken either
the one in the quartet model or in the doublet model.
The Ohmic behavior $r(\omega ) \propto  \omega $ at $\omega \sim 0$ is
then a consequence of the presence of a gap;
\( \:
\Omega _{0} > 0 \,.
\: \)

Departure from the Ohmic behavior appears at some frequency scale and
if this scale also coincides with the cutoff scale of the high frequency 
contribution, one may regard the model as an extended Ohmic model.
On the other hand, models that contain several distinct frequency scales 
may considerably differ from the Ohmic model.
Our main focus in the response weight in the present paper
is on the Ohmic model modified at high frequencies: extended Ohmic
models.

\vspace{0.5cm} 
{\bf \lromn 3 B. \hspace{0.3cm}
Field theory model and perturbative analysis
}

\vspace{0.5cm} 
The infinitely many environment variable $Q_{a}$
can actually be a composite field
of more fundamental fields. In order to discuss a fully relativistic
field theory, we also consider the case in which the system variable $q_{k}$
is a Fourier component of some scalar field $\varphi (x)$ put in
a homogeneous medium.
As a simplest field theory model of non-linear system-environment interaction,
we thus take a Yukawa type of interaction with a new bose field $\chi (x)$;
\begin{equation}
L_{{\rm int}} = \int_{V}\,d^{3}x\,{\cal L}_{{\rm int}} \,, \hspace{0.5cm} 
{\cal L}_{{\rm int}} = -\,\frac{\mu }{2}\,\varphi \chi ^{2}
\,,
\end{equation}
with $\mu $ a parameter of mass dimension. 
$V$ is the normalization volume.
By identifying the Fourier-mode with
\begin{equation}
q_{k}\,e^{i\vec{k}\cdot \vec{x}} = 
\frac{1}{\sqrt{2\omega _{k}\,V}\,} \;(\,a_{k} + a^{\dag }_{-k}\,)\,
e^{i\vec{k}\cdot \vec{x}} \,, 
\end{equation}
we deduce the environment variable coupled to it,
\begin{equation}
\sum_{a}\,c_{a}Q_{a} =
\frac{\mu }{2}\,\int_{V}\,d^{3}x\,\chi ^{2}(x)\,e^{-\,i\vec{k}\cdot \vec{x}}
\,. 
\end{equation}
Henceforth we regard the field $\chi $ as a fundamental variable
of the environment, assumed to be in thermal equilibrium.

The correlation function $\alpha (\tau )$, or more conveniently its
Fourier transform $\alpha (\omega )$, can be computed as in the real
time formalism at finite temperatures. To lowest order in $\mu ^{2}$
\begin{equation}
i\,\alpha_{k} (\tau - s)  = i\,(\frac{\mu }{2})^{2}\,
\int_{V}\,d^{3}x\,e^{i\vec{k}\cdot \vec{x}}\;
{\rm tr}\;
\left( \,T\,[\,\chi ^{2}(\vec{x}\,, \tau )\,
\chi ^{2}(\vec{0}\,, s)\,]\,\rho_{\beta } 
\,\right) 
\end{equation}
is calculable in terms of the $\chi-$propagator in the momentum space;
\begin{equation}
i\,\left( \,\frac{1}{\omega ^{2} - \vec{k}^{2} - m_{\chi }^{2} +
i\epsilon } - \frac{2\pi i}{e^{\beta \omega _{k}} - 1}\,
\delta (\omega ^{2} - \omega _{k}^{2})\,\right) \,.
\end{equation}
This is of course identical to the Fourier transform of the one we
are already familiar with, $\alpha _{R}(\tau ) + 
i\alpha _{I}(\tau )$.
We omit the suffix $\chi $ for the $\chi-$mass $m_{\chi }$ in what
follows.

As in the $T=0$ field theory it is easiest to compute
the discontinuity instead of the full diagramatic contribution.
Physically the most transparent way 
to compute the imaginary part of the self-energy diagram,
\( \:
\Im \,\Pi (\omega ) \,, 
\: \)
is to use the analytically continued expression  \cite{weldon}
from the imaginary-time formalism, as outlined in the preceding section. 
The analytically continued imaginary part, or more precisely the
discontinuity, directly gives the response weight;
\begin{equation}
r (\omega ) = 2\Im \,\Pi (\omega ) = 2\omega\, \Gamma (\omega )
\,, \hspace{0.5cm} 
\Im \,\Pi (\omega ) \equiv  \frac{1}{2i}\,\left( \,
\Pi (\omega - i\epsilon ) - \Pi (\omega + i\epsilon )\,\right) \,,
\end{equation}
where $\Gamma (\omega )$ is interpreted as a decay rate in
thermal medium.

In the subthreshold region of 
\( \:
|\omega | < k
\: \)
one loop contribution is given by
\begin{eqnarray}
\Im \Pi\, (\omega ) &=& \frac{\mu ^{2}}{16\pi k}\,
\int_{- \omega _{-}}^{\infty }\,dE\,\left( \,n(E) - n(E + \omega )\,\right)
\,, \label{subthreshold imaginary part} \\
\omega _{\pm } &=&
\frac{\omega }{2} \pm \frac{k}{2}\,
\sqrt{\,1 - \frac{4m^{2}}{\omega ^{2}-k^{2}}\,} \,,
\end{eqnarray}
where
\begin{equation}
n(E) = \frac{1}{e^{\beta E} - 1}
\end{equation}
is the Planck distribution function.
With
\begin{equation}
n(E) - n(E + \omega ) = n(E)(\,1 + n(E + \omega )\,) -
n(E + \omega )(\,1 + n(E)\,) \,, 
\end{equation}
the imaginary part (\ref{subthreshold imaginary part}) for
$|\omega | < k$ is a sum of the two contributions,
\( \:
\chi + \varphi \rightarrow \chi 
\: \)
and its inverse processes that occur in thermal medium.
The factor $1 + n$ represents the effect of stimulated boson emission.
Near $\omega = 0$,
\begin{equation}
r(\omega ) \sim 
\frac{\mu ^{2}}{8\pi }\,\frac{\omega }{k}\,n(E_{*}) \,, \hspace{0.5cm} 
E_{*} = \frac{1}{2}\, \sqrt{\,k^{2} + 4m^{2}\,} \,.
\end{equation}
This in turn gives the friction coefficient,
\begin{equation}
\eta = \pi \,r'(0) = 
\frac{\mu ^{2}}{8\,k}\,\,n(E_{*}) \,.
\label{friction in field th} 
\end{equation}
As emphasized in \cite{weldon}, the decay rate
\begin{equation}
\Gamma (0) = \frac{1}{2}\, r'(0) = \frac{\eta }{2\pi } 
\end{equation}
can be interpreted as the relaxation rate of thermal disturbance of
the $\varphi-$field when departure from equilibrium is introduced.
At high and low temperatures the friction thus computed behaves as
\begin{equation}
\eta \rightarrow \frac{\mu ^{2}\,T}{4k\,\sqrt{\,k^{2} + 4m^{2}\,}}
\,, \; (T \rightarrow \infty ) \,, 
\hspace{0.5cm} 
\eta \rightarrow \frac{\mu ^{2}}{8k}\,e^{-\,\sqrt{\,k^{2} + 4m^{2}
\,}\,/(2T)} \,, \; (T \rightarrow 0) \,.
\end{equation}

On the other hand, for $\omega > \sqrt{\,k^{2} + 4m^{2}\,}$ 
relevant physical processes
are 
\( \:
\varphi \:\leftrightarrow  \: \chi + \chi \,.
\: \)
Since
\begin{equation}
(\,1 + n(E)\,)(\,1 + n(E + \omega )\,)- n(E)n(E + \omega ) = 
1 + n(E) + n(E + \omega ) \,,
\end{equation}
the imaginary part of the self-energy diagram is given by
\begin{eqnarray}
\Im \Pi\, (\omega ) &=& \frac{\mu ^{2}}{32\pi k}\,
\int_{\omega _{-}}^{\omega _{+}}\,dE\,\left( \,1 + 2n(E)\,\right)\,,  \\
&\rightarrow &
\frac{\mu ^{2}}{32\pi }\,\sqrt{\,1 - \frac{4m^{2}}{\omega ^{2} - k^{2}}\,}
\,(\,1 + 2e^{-\,\beta \omega /2}\,) \,, 
\end{eqnarray}
as $\omega  \rightarrow \infty $. 
When divided by $\omega $, the first term here 
\( \:
\Gamma (\omega ) = \Im \,\Pi (\omega )/\omega 
\: \)
is 
\( \:
\approx \mu ^{2}/(32\pi \,\omega ) \,, 
\: \)
which is the decay rate for $\varphi \rightarrow \chi \chi $ in
vacuum, including the effect of prolonged lifetime at
\( \:
\omega  \gg m_{\varphi } \,.
\: \)
The response weight given by the imaginary part of the self-energy at one loop
is shown in fig.1 for a set of the parameters 
\( \:
(\,k \,, \; m \,, \; T \,) \;
\: \), 
both in the sub- and super-threshold regions.
Also shown here is an approximation to this weight, 
eq.(\ref{approximate field th model}) given below.

\vspace{0.5cm} 
{\bf \lromn 3 C. \hspace{0.3cm}
High frequency cutoff and extended Ohmic model
}

\vspace{0.5cm} 
The truncated form of the response weight is often adequate in discussion
of most physical quantities, especially at high temperatures.
But the frequency integral for the correlation $\alpha _{R}(\tau )$
is not convergent if $r(\omega  ) \propto \omega $ all the way up to
$\omega = \infty $. Actually there exists a high frequency cutoff;
\begin{eqnarray}
r^{(4)}(\omega ) &\rightarrow& \frac{\eta \Omega ^{4}}{\pi }\,
\frac{1}{\omega ^{3}} \,, \hspace{0.5cm} {\rm for \; quartet \; model}
 \,, \\
r^{(\beta )}(\omega ) &\rightarrow& 
\frac{\mu ^{2}}{8\pi }\,e^{-\,\beta \omega /2} \,, 
\hspace{0.5cm} {\rm for \; field \; theory \; model} \,,
\end{eqnarray}
assuming $\Omega \gg \gamma $ for the quartet model.
We subtracted the $T = 0$ contribution in writing the limiting form
$r^{(\beta )}(\omega )$ of the field theory model.

The cutoff frequency appears quite different in the two cases,
at $\Omega $ in the quartet model and at $T$ in the field theory model
at finite temperatures.
Moreover, the high frequency tail is also different and we 
observe a very fast convergence of the temperature dependent term
in the field theory model.
In the literature the straightforward cutoff at some $\omega = \Omega $
is also used frequently \cite{caldeira-leggett 83}, \cite{unruh-zurek}.
It is important, however, to note that 
there may be temperature dependence of the cutoff $\Omega $.
We believe that in some realistic applications the relation
$\Omega \propto T$ holds at high temperatures. 
For instance, if one computes excitation energy
in the form of mass correction in finite temperature field theory,
one usually finds that
\( \:
\Omega \approx g\,T \, 
\: \) 
at high temperatures, with $g$ some dimensionless coupling constant.
In the rest of discussion we assume this proportionality at high
temperatures, whenever we need realistic estimates.
At very low temperatures this relation does not hold, and the cutoff
$\Omega $ reflects some intrinsic scale of the environment, to which
we have to pay attention.

It appears that the constant weight at a large $\omega $,
\( \:
r(\omega ) \sim \mu ^{2}/(16\pi ) \,, 
\: \)
in the field theory model must be retained at low temperatures.
This constant term does not contribute to the imaginary part
of the correlation $\alpha _{I}(\tau )$ 
(see eq.(\ref{imaginary correlation})). We may thus use the subtracted
weight,
\( \:
r(\omega ) - r(\infty )
\: \)
for $\alpha _{I}(\tau )$, which behaves asymptotically like
$O[1/\omega ^{2}]$ as $\omega \rightarrow \infty $. 
But if the high frequency contribution of the response weight with
\( \:
r(\omega ) \rightarrow \mu ^{2}/(16\pi )
\: \)
dominates over the low frequency contribution
in the $\omega $ integral of the real part of the correlation,
it does not fit into the approximation scheme of the localized friction,
at least in the present simplest form.
If we do not deal with this case and remain in the realm of the localized
friction approximation, the field theory model belongs to an extended
Ohmic model with a cutoff
\( \:
\Omega \approx \sqrt{\,k^{2} + 4m_{\chi }^{2}\,}
\: \). 
Physically this means that
we identify the main process of dissipation in the field theory model as 
\( \:
\chi + \varphi \leftrightarrow \chi 
\: \)
instead of 
\( \:
\varphi \rightarrow \chi + \chi \,.
\: \)

A better approximation of the field theory model is to replace the one
loop response weight by a continuous function that well reproduces
the correct behavior  both in the
$\omega = 0$ and in the $\omega =\infty $ regions. If one ignores a mismatch
in the threshold region,
\( \:
 \omega = k \sim  \sqrt{\,k^{2} + 4m_{\chi }^{2}\,}  \,,
\: \)
then an example is given by
\begin{eqnarray}
r_{A}(\omega ) &=& \frac{\mu ^{2}}{16\pi }\,
\frac{2\omega }{2\omega + N_{0}\,k}\,
\coth \left( \,\frac{\omega + \sqrt{\,k^{2} + 4m^{2}\,}}{4T} \,\right)
\,, 
\label{approximate field th model} \\
N_{0} &=& e^{\sqrt{\,k^{2} + 4m^{2}\,}/(2T) } + 1 \,.
\end{eqnarray}
As illustrated in fig.1, this gives a reasonable description of the
response weight except in the threshold region.
The threshold region alone may be described by a quartet type of model,
but we shall not elaborate on this problem any further.

Thus in all models discussed in this section the response weight 
has the Ohmic behavior at low frequencies. 
Differences in the models are in the form of the cutoff function 
$f(\frac{\omega }{\Omega })$ and the cutoff scale $\Omega $;
\( \:
f(x) \propto 1/x^{4}
\: \)
in the quartet model,
\( \:
f(x) \propto 1/x^{2}
\: \)
in the doublet model, and
\( \:
f(x) \propto 1/x
\: \)
in the one-loop field theory model.
If one focuses on the late time behavior, the form of the cutoff function
is not critical unless the environment temperature
\( \:
T \ll \Omega \,.
\: \)
In subsequent applications we use a simpler cutoff of the form,
\( \:
e^{-\,\omega /\Omega } \,,
\: \)
whenever necessary in analytic estimates.
On the other hand,
for numerical computation we cutoff various $\omega $ integrals at $\Omega $ 
as a simple means of computation.
This should be sufficient practically when application is limited to
high temperatures. 
Comparison with other forms of the cutoff has been attempted in a few
sample cases.
We first checked that results of numerical computation shown below
are insensitive to the value of the straightforward cutoff scale 
$\Omega $ beyond some value, $\Omega \geq 50$
in our unit. Use of the quartet model that has the same friction
constant  and the same effective maximum value of the response weight 
as in the case of the cutoff method
neither did  produce any recognizable change of our results.

\vspace{1cm}
\section{Time dependent harmonic oscillator in thermal bath}

\vspace{0.5cm} 
\hspace*{0.5cm} 
We have so far discussed how to characterize the environment using
the influence functional. 
We now apply this formalism to the dynamical system governed by
the Lagrangian of time dependent harmonic oscillator;
\begin{equation}
L_{q}
= \frac{1}{2}\, \dot{q}^{2} - \frac{1}{2}\, \omega ^{2}(t)q^{2} \,.
\end{equation}
The main purpose to treat this class of models
is to clarify effect of the environment
on the parametric particle production. Besides this direct goal
this class of models can cope with many interesting cases of application,
some of which we hope to cover in subsequent works.
We therefore explain how to derive basic formulas somewhat in detail.

The basic quantity we focus on is the path integral over the system
variable $q$;
\begin{eqnarray}
&&
\int\,{\cal D}q(\tau )\,\int\,{\cal D}q'(\tau )\,
\exp \left( \, iS[q(\tau )] - iS[q'(\tau )] \,\right) \,
{\cal F}[\,q(\tau )\,, q'(\tau )\,]  \nonumber \\
&& \hspace*{0.5cm} 
=
N\,\int\,{\cal D}\xi (\tau )\,\int\,{\cal D}X(\tau )\,
\exp \left[\,
\frac{i}{2}\,\int_{0}^{t}\,d\tau 
\left( \,\dot{\xi }(\tau )\,\dot{X}(\tau ) - 
\omega ^{2}(\tau )\,\xi (\tau )X(\tau )\,\,\right) \right.
\nonumber \\
&& 
\hspace*{0.5cm} \left.
-\, \int_{0}^{t }\,d\tau \,\int_{0}^{\tau }\,ds\,
\left( \,\xi (\tau )\alpha_{R}(\tau - s)\xi (s) + i\xi (\tau )
\alpha _{I}(\tau - s)X(s)\,\right)
\,\right] \,,
\end{eqnarray}
with 
\( \:
\xi (\tau ) = q(\tau ) - q'(\tau ) \,, \hspace{0.3cm} 
X(\tau ) = q(\tau ) + q'(\tau ) \,.
\: \)
The effect of the correlation $\alpha (\tau - s)$ introduces
non-local interactions among the system variable $q$, 
in general difficult to deal with.
Even diagrammatic expansion may not work with the presence of time dependence
in $\omega ^{2}(\tau ) $.
But the most important cases can be discussed using the approximate
localized friction for $\alpha _{I}(\tau - s)$, to which we shall turn.
In this case one can explicitly perform the path integral in terms of
a classical solution for $q(\tau )$.
The problem is then reduced to analysis of a rather simple
classical system.

\vspace{0.5cm} 
{\bf \lromn 4 A. \hspace{0.3cm}
Localized friction}

\vspace{0.5cm} 
The notion of localized friction arises by focusing on the late time
behavior of $\alpha _{I}(\tau )$. By the late time here
we mean the asymptotic
time that is much larger than any time scale of excitation in the
environment; $t \gg 1/$ (maximal excitation frequency) $\approx 1/\Omega $. 
The limiting form is then given by
\begin{eqnarray}
&&
\alpha _{I}(\tau )  = -\,\int_{0}^{\infty }\,d\omega 
\,r(\omega )\,\sin (\omega \tau ) \:\rightarrow  \: 
\eta \,\delta '(\tau ) \,, \\
&&
\eta = -\,2\int_{0}^{\infty }\,d\tau \,\tau\,\alpha _{I}(\tau )
= \pi \,r'(0) \,, \hspace{0.5cm} {\rm for}\; 
r(\omega = \infty ) = 0 \,, \\
&&
\alpha _{R}(\tau ) \:\rightarrow  \: \frac{\eta }{\pi }\,
\int_{0}^{\infty }\,
d\omega \,\omega \,f(\frac{\omega }{\Omega })\,
\coth (\frac{\beta \omega }{2})\,\cos (\omega \tau )
\,.
\end{eqnarray}
The essence of this approximation is the low frequency truncation to
the response weight,
\begin{equation}
r(\omega ) = \omega \,r'(0) = \frac{\eta }{\pi }\,\omega \,,
\end{equation}
such that
\( \:
f(x) \approx 1 \,.
\: \)
We already computed in the preceding section 
the friction constant $\eta $ in a variety of models,
specifically in the quartet model and in the field theory model.

Although the dissipation given by $\alpha _{I}(\tau )$ is local by the
time scale of $1/\Omega $, the damping time scale of the noise kernel 
$\alpha _{R}(\tau )$ may differ. It can be shown that at high temperatures
of $T\geq \Omega $ the locality of the correlation holds for
$\tau \geq 1/\Omega $, giving
\begin{equation}
\alpha _{R}(\tau ) \sim \frac{\eta T}{\pi }\,\delta (\tau ) \,.
\end{equation}
On the other hand, at low temperatures of $T \leq \Omega $
the local approximation may be inadequate and a high frequency cutoff is
needed. As already discussed, we replace in this case 
the response weight by the more appropriate one; for instance 
in the quartet model
\begin{equation}
r(\omega ) = 
\frac{\eta }{\pi }\,\omega \,f(\frac{\omega }{\Omega })
\,, \hspace{0.5cm} 
f(x) = \frac{1}{(\,x^{2} - 1 + \gamma ^{2}/(4\Omega ^{2})\,)^{2} 
+ \gamma ^{2}/\Omega ^{2}}
\,, 
\end{equation}
or by the simple frequency cutoff at $\omega = \Omega $ with
$f(x) = e^{-\,x}$, as is often practiced.

With $\alpha _{I}(\tau ) \:\propto  \: \delta '(\tau )$, the exponent
factor in the path integral becomes, including the mass renormalization
effect,
\begin{eqnarray}
&&
-\,
\frac{i}{2}\,\int_{0}^{t}\,d\tau \,
X(\tau )\,(\,\frac{d^{2}}{d\tau ^{2}}+ \omega _{R}^{2}
(\tau )\,)\,\xi(\tau )  + \frac{i}{2}\,\eta \,\int_{0}^{t}\,d\tau \,
X(\tau )\,\dot{\xi }(\tau ) 
\nonumber \\
&& \hspace*{0.5cm} 
+ \frac{i}{2}\,\left[\,X\dot{\xi }\,\right]^{t}_{0}
- \frac{i}{2}\,\eta \,\xi _{f}X_{f} \,{\rm sign}(t) 
- \int_{0}^{t }\,d\tau \,\int_{0}^{\tau }\,ds\,
\xi (\tau )\alpha_{R}(\tau - s)\xi (s) \,.
\end{eqnarray}
A remarkable feature of this formula is that  a part of the action 
excluding the $\alpha _{R}$ term is 
a local integral described in terms of a renormalized frequency 
$\omega _{R}(\tau )$ and the friction term $\propto \eta $.
This much is enough to considerably simplify calculation of the transition
amplitude of the $q-$system.

Path integral over the system variable $q$ is standard; it can
be decomposed into contributions from the classical path and from a
Gaussian fluctuation around this classical path.
Let us first write down the equation for the classical path of
$\xi (\tau )$, which is obtained by functional differentiation
with respect to $X(\tau )$;
\begin{eqnarray}
\left( \,\frac{d^{2}}{d\tau ^{2}} + \omega _{R}^{2}(\tau ) - 
\eta \,\frac{d}{d\tau }\,\right)\,\xi (\tau ) = 0 \,.
\end{eqnarray}
This equation is best analyzed by introducing a new $y(\tau )$,
\begin{eqnarray}
&&
y(\tau ) \equiv e^{\eta \tau /2}\,\xi (\tau ) \,, \\
&&
\left( \,\frac{d^{2}}{d\tau ^{2}} + \omega _{R}^{2}(\tau ) - 
\frac{\eta ^{2}}{4}\,\right)\,y(\tau ) = 0 \,.
\end{eqnarray}
It is convenient to use two types of solution $u(\tau )$ and $v(\tau )$
with different boundary conditions;
\begin{equation}
u(0) = 0 \,, \hspace{0.5cm}  v(t) = 0 \,.
\end{equation}
Actually if $u(\tau )$ is known, the other is given by
\begin{equation}
v(\tau ) = c\,u(\tau )\,\int_{\tau }^{t}\,\frac{d\tau '}{u^{2}(\tau ')}
\,,
\end{equation}
with $c$ a constant to be determined by normalization.

The normalization of these solutions is fixed by specifying the value of
the constant Wronskian,
\begin{equation}
u(\tau )\dot{v}(\tau ) - v(\tau )\dot{u}(\tau ) = -\,\dot{u}(0)v(0) =
u(t)\dot{v}(t) \,.
\end{equation}
For instance, by setting this to the initial shifted frequency
\begin{equation}
\tilde{\omega }_{0} \equiv \sqrt{\,\omega _{0}^{2} - \frac{\eta ^{2}}{4}\,}
\,, \hspace{0.5cm} 
\omega _{0}^{2} \equiv \omega _{R}^{2}(0)
\,, 
\end{equation}
one may have, consistently with the Wronskian relation,
\begin{equation}
\dot{u}(0) = \tilde{\omega }_{0} \,, \hspace{0.5cm} 
v(0) = 1 \,. \label{normalization} 
\end{equation}
In the rest of this paper we leave formulas as general as possible
without specifying the normalization condition,
but in places where we specifically 
need normalization constants we shall use this normalization.

The classical $\xi $ path is then
\begin{equation}
\xi _{{\rm cl}}(\tau ) = 
\xi _{f}\,\frac{u(\tau )}{u(t)}\,e^{-\,\eta (t - \tau )/2}
+ \xi _{i}\,\frac{v(\tau )}{v(0)}\,e^{\eta \tau /2} \,.
\end{equation}
Fortunately, we do not need details of the other solution 
$X_{{\rm cl}}(\tau )$; 
only the boundary values $X_{i} \,, X_{f}$ are needed for the $X$ part.

The classical action is thus computed after getting
\( \:
\dot{\xi }_{i} \,, \dot{\xi }_{f}
\: \)
from $\xi _{{\rm cl}}(\tau )$,
\begin{eqnarray}
i\,S_{{\rm cl}} &=& -\,\frac{U}{2}\,\xi _{f}^{2} - \frac{V}{2}\,\xi _{i}^{2}
- W\,\xi _{i}\,\xi _{f} - \frac{i}{2}\,\eta \,\xi _{f}X_{f} \nonumber \\
&+& 
\frac{i}{2}\,X_{f}\,
\left( \,\xi _{f}\,(\,\frac{\eta }{2} + \frac{\dot{u}(t)}{u(t)}\,) +
\xi _{i}\,\frac{\dot{v}(t)}{v(0)}\,e^{\eta t/2}\,\right) \nonumber \\
&-&
\frac{i}{2}\,X_{i}\,
\left( \,\xi _{i}\,(\,\frac{\eta }{2} + \frac{\dot{v}(0)}{v(0)}\,) +
\xi _{f}\,\frac{\dot{u}(0)}{u(t)}\,e^{-\eta t/2}\,\right) \,,\\
U &=& 2\,\int_{0}^{t }\,d\tau \,\int_{0}^{\tau }\,ds\,
\frac{u(\tau )u(s)}{u^{2}(t)}\,\alpha _{R}(\tau - s)\,
e^{-\,\eta \,(2t - \tau - s)/2} \,, 
\\
V &=& 2\,\int_{0}^{t }\,d\tau \,\int_{0}^{\tau }\,ds\,
\frac{v(\tau )v(s)}{v^{2}(0)}\,\,\alpha _{R}(\tau - s)\,
e^{\eta \,(\tau + s)/2} \,, 
\\
W &=& \int_{0}^{t }\,d\tau \,\int_{0}^{\tau }\,ds\,
\frac{u(\tau )v(s) + u(s)v(\tau )}{v(0)u(t)}\,\alpha _{R}(\tau - s)\,
e^{-\,\eta \,(t - \tau - s)/2} \,.
\end{eqnarray}

The second step of computation on the Gaussian fluctuation around
the classical path is trivial for 
\( \:
\delta X(\tau ) = X(\tau ) - X_{{\rm cl}}(\tau ) \,, 
\: \)
because the action is linear in this variable. This simply gives 
a delta-functional containing a linear term in $\delta \xi (\tau )$;
\begin{equation}
\delta \left( \, (\,
\frac{d^{2}}{d\tau ^{2}}+ \omega _{R}^{2}(\tau )- 
\eta \,\frac{d}{d\tau }
\,)\,\delta \xi(\tau )\,\right) \,.
\end{equation}
Since the boundary condition is 
\( \:
\delta \xi (0) = \delta \xi (t) = 0 \,, 
\: \)
the only solution that makes the argument vanish is the trivial one,
\( \:
\delta \xi (\tau ) = 0 \,.
\: \)
This makes the rest of $\delta \xi (\tau )$ integration also trivial.
The net result is then a determinant factor,
\begin{equation}
\left( \,{\rm det}\; [\,
\frac{d^{2}}{d\tau ^{2}}+ \omega _{R}^{2}(\tau )- \eta \,\frac{d}{d\tau }
\,]\,\right)^{-1} \,.
\end{equation}
The Feynman-Vernon formula (\ref{influence-f def})  
respects unitarity to arbitrary orders of
the correlation $\alpha (\tau )$. This makes it possible to 
handle the determinant factor above by a total normalization factor $N(t)$,
which is fixed in the end of calculation by the condition of the density
matrix,
\( \:
{\rm tr}\; \rho ^{(R)} = 1
\: \).

Finally, we note that although non-local dissipation is beyond the scope
of the present work, there are physical situations that inevitably go
beyond the local friction approximation. 
In the extended Ohmic model that has a physical high
frequency cutoff $\Omega $, the short time behavior at
$\tau \leq 1/\Omega $ cannot be trusted, because the picture of the
local friction given by
\( \:
\alpha _{I}(\tau - s) \:\propto  \: \delta '(\tau - s)
\: \)
fails. 
If the time scale of the system change is $\omega _{0}$, it must be
that $\omega _{0} \leq \Omega $. Otherwise what takes place in the time
scale of $\tau \leq 1/\omega _{0}$ cannot be trusted.
The non-local dissipation becomes important in the case of
\( \:
\omega _{0} > \Omega \,.
\: \)

\vspace{0.5cm} 
{\bf \lromn 4 B. \hspace{0.3cm}
Reduced density matrix, energy distribution, and entropy}

\vspace{0.5cm} 
A useful quantity for determining the behavior of the quantum system
under the environmental interaction is the reduced density matrix
$\rho ^{(R)}$ obtained by convoluting with the initial state of the $q-$system;
\begin{eqnarray}
&&
\rho ^{(R)}(X_{f}\,, \xi _{f}) = 
N(t)\,\int\,{\cal D}\xi (\tau )\,\int\,{\cal D}X(\tau )\,
\int\,d\xi _{i}\,\int\,dX_{i}\,
\nonumber \\
&& \hspace*{0.5cm} 
\cdot \rho_{0}(\xi _{i} \,, X_{i})\,
\exp \left( \, iS[q(\tau )] - iS[q'(\tau )] \,\right)
\,{\cal F}[\,q(\tau )\,, q'(\tau )\,] 
\,.
\end{eqnarray}
We take for the density matrix of the initial $q-$state,
\begin{equation}
\hspace*{-1cm}
\rho_{0} (\xi _{i} \,, X_{i}) = 
\sqrt{\,\frac{\omega _{0}}{\pi \,\coth (\beta _{0}\omega _{0}/2)}\,}\,
\exp [\,- \,\frac{\omega _{0}}{4}\,\coth (\frac{\beta _{0}\omega _{0}}{2})\,
\xi _{i}^{2} - \frac{\omega _{0}}{4}\,\tanh(\frac{\beta _{0}\omega _{0}}{2})
\,X_{i}^{2}\,]  \,,
\end{equation}
which gives a thermal state at temperature $T_{0} = 1/\beta _{0}$.
We use this initial $\xi-$state as a convenient means of computation,
since we can treat both the pure ground state (in the limit of 
$T_{0} \rightarrow 0$) and a thermal state.

A straightforward Gaussian integration then gives the reduced density
matrix,
\begin{eqnarray}
\rho ^{(R)}(X_{f}\,, \xi _{f})  
&=& \frac{N(t)}{\sqrt{2\pi }}\,e^{-\,\eta  t}\,K^{-1/2}\,
\cdot 
\exp [\,- {\cal A}\,X_{f}^{2} - {\cal B}\,\xi _{f}^{2} 
+ i\,{\cal C}\,X_{f}\xi _{f}\,] \,, \\
{\cal A} &=& \frac{1}{8}\,(\frac{\dot{u}(0)}{u(t)})^{2}\,K^{-\,1} \,, \\
{\cal B} &=& \frac{A}{2} + 
\frac{1}{4\omega _{0}}\,\coth (\frac{\beta _{0}\omega _{0}}{2})\,
(\frac{\dot{u}(0)}{u(t)})^{2}\,e^{-\eta t} - \frac{L^{2}}{2\,K} \,, \\
{\cal C} &=& \frac{1}{2}\,\frac{\dot{u}(0)}{u(t)}\,\frac{L}{K}
- \frac{1}{2}\,\left(\,\frac{\eta }{2} - \frac{\dot{u}(t)}{u(t)}\,\right)
\,, \\
K &=& B + e^{-\,\eta t}\,
\frac{\omega _{0}}{2}\,\coth  (\frac{\beta _{0}\omega _{0}}{2})
\left( \, 1 + \frac{1}{\omega _{0}^{2}}\,
(\,\frac{\eta }{2} + \frac{\dot{v}(0)}{v(0)}\,)^{2}
\,\right) \,, \label{k-equation} \\
L& =& 
C + \frac{e^{-\,\eta t}}{2\omega _{0}}\,
\coth (\frac{\beta _{0}\omega _{0}}{2})\,
\frac{\dot{u}(0)}{u(t)}\,(\frac{\eta }{2} +  \frac{\dot{v}(0)}{v(0)})
\,,
\\
(\,A \,, B \,, C\,) 
&=& \int_{0}^{\infty }\,d\omega \,\coth (\frac{\beta \omega }{2})\,
r(\omega )\,(a\,, b\,, c)(\omega )  \,, \\
a(\omega ) &=&
\left|\,\int_{0}^{t}\,ds\,\frac{u(s)}{u(t)}\,
e^{i\omega s - \frac{\eta }{2}(t - s)}\,\right|^{2} \,, 
\\ 
b(\omega ) &=& 
\left|\,\int_{0}^{t}\,ds\,\frac{v(s)}{v(0)}\,
e^{i\omega s - \frac{\eta }{2}(t - s)}\,\right|^{2} 
\,, \\ 
c(\omega ) &=& \Re \,\left( \,
\int_{0}^{t}\,ds\,\frac{u(s)}{u(t)}\,
e^{i\omega s - \frac{\eta }{2}(t - s)}\cdot 
\int_{0}^{t}\,ds\,\frac{v(s)}{v(0)}\,
e^{-i\omega s - \frac{\eta }{2}(t - s)}\,\right) \,.
\label{last c-eq}
\end{eqnarray}
The normalization is determined from probability conservation,
\begin{eqnarray}
\int_{-\infty }^{\infty }\,dq\,\rho ^{(R)}(2q \,, 0) = 1 \,, 
\end{eqnarray}
leading to
\begin{eqnarray}
\rho ^{(R)}(X_{f}\,, \xi _{f}) = 2\,\sqrt{\frac{{\cal A}}{\pi }}\,
\exp [\,-\,{\cal A}\,X_{f}^{2} - {\cal B}\,\xi _{f}^{2} +
i\,{\cal C}\,X_{f}\,\xi _{f}\,] \,.
\label{density matrix} 
\end{eqnarray}
This is a fundamental result of the present work:
The reduced density matrix is determined by two factors:
a single classical solution
$u(\tau )$ (and related $v(\tau )$) for the system, and the response
weight $r(\omega )$ for the system-environment interaction.
The only approximation we need to derive this formula is
the local form of $\alpha _{I}(\tau - s)$.

The diagonal part of the reduced density matrix with $q = q'$, namely
with $\xi = 0$, is
\begin{eqnarray}
\rho ^{(R)}(X_{f} = 2q \,, 0) = 2\,\sqrt{\frac{{\cal A}}{\pi }}\,
\exp [\,-\,4{\cal A}\,q^{2} \,] \,.
\end{eqnarray}
The width of the Gaussian peak of $\approx 1/\sqrt{{\cal A}}$ is a measure of
how the system behaves, but we know more details in the form of
the complete density matrix.

It is sometimes useful to transform the density matrix in the
configuration space to the Wigner function $f_{W}(x\,, p)$,
\begin{eqnarray}
f_{W}(x \,, p) & \equiv& \int_{-\infty }^{\infty }\,d\xi \,
\rho^{(R)}(q = x + \frac{\xi }{2} \,, q' = x - \frac{\xi }{2})\,
e^{-\,i\,p \xi }
 \,, \\
&=&
\sqrt{\frac{4{\cal A}}{{\cal B}}}\,\exp [\,-\,4{\cal A}\,x^{2}
- \frac{(p - 2{\cal C}\,x)^{2}}{4{\cal B}}\,] \,.
\end{eqnarray}
Expectation value of the number operator, namely the occupation number, 
in terms of the reference frequency,
taken to be the initial one $\omega _{0}$,
is calculated most easily from the Wigner function;
\begin{equation}
\langle N \rangle \equiv  
\langle \,-\,\frac{1}{2\omega _{0} }\,\frac{d^{2}}{dq^{2}} +
\frac{\omega _{0}}{2}\,q^{2} - \frac{1}{2}\, \rangle 
= \frac{{\cal B}}{\omega _{0}}
 + \frac{{\cal C}^{2}}{4\omega _{0}\,{\cal A}}
 + \frac{\omega _{0}}{16{\cal A}} - \frac{1}{2} \,,
\end{equation}
which consists, except the trivial $\frac{1}{2}$, 
of two terms, the term $\omega _{0}/(16\,{\cal A})$ from the Gaussian
width of the diagonal density matrix element and the rest from
the kinetic term $-\,\frac{d^{2}}{dq^{2}}$.
We shall discuss the rate of particle production in more detail
when we turn to specific applications, namely the two cases 
of constant and periodic frequency $\omega ^{2}(\tau )$.

One can work out the frequency distribution of populated harmonic oscillator
levels. Suppose that one asks the probability of finding the system in
the unit energy interval around
\( \:
\omega = \frac{p^{2}}{2} + \frac{\omega _{0}^{2}}{2}\,x^{2} \,.
\: \)
We get the energy distribution function $f(\omega )$ from the Wigner function;
\begin{eqnarray}
f(\omega ) &\equiv& \int\,\frac{dx\,dp}{2\pi }\,f_{W}(x\,, p)\,
\delta (\,\omega - \frac{p^{2}}{2} - \frac{\omega _{0}^{2}}{2}\,x^{2}\,)
\\
&=&
\frac{2}{\omega _{0}}\,\sqrt{\frac{{\cal A}}{{\cal B}}}\,
\exp \left( \,-\,\frac{\omega}{\omega _{0}}\, 
(\,\frac{4{\cal A}}{\omega _{0}} + \frac{4{\cal C}^{2} + \omega _{0}^{2}}
{4\omega _{0}\,{\cal B}}\,)\,\right)\,I_{0}(\,\frac{\omega }{\omega _{0}}
\,z\,) \,, \label{energy distribution} \\
z &=& \left( \,(\,\frac{4{\cal A}}{\omega _{0}} 
+ \frac{4{\cal C}^{2} + \omega _{0}^{2}}
{4\omega _{0}\,{\cal B}}\,)^{2} - \frac{4{\cal A}}{{\cal B}}\,\right)^{1/2}
\,.
\end{eqnarray}
Here 
\begin{equation}
I_{0}(w) = \frac{1}{\pi }\,\int_{0}^{\pi }\,d\theta \,e^{-\,w\cos \theta }
\end{equation}
is the modified Bessel function, having 
\begin{equation}
I_{0}(w) \:\rightarrow  \: \frac{1}{\sqrt{\,2\pi w\,}}\,e^{w} 
\hspace{0.3cm} ({\rm as \;} w \rightarrow \infty ) \,, \hspace{0.5cm} 
\:\rightarrow  \: 1 \hspace{0.3cm} ({\rm as \;} w \rightarrow 0) \,.
\end{equation}

This distribution differs from the thermal one and has some
interesting features.
At $\omega \rightarrow \infty $, it decreases as
\begin{eqnarray}
f(\omega ) &\rightarrow& \frac{2}{\sqrt{\,2\pi \,\omega _{0}\omega \,z\,}}
\,\sqrt{\,\frac{{\cal A}}{{\cal B}}\,}\,e^{-\,\omega \,D}\,, 
\\
D &=& \frac{1}{\omega _{0}}\,
\left[\,
\frac{4{\cal A}}{\omega _{0}} + \frac{4{\cal C}^{2} + 
\omega _{0}^{2}}{4\omega _{0}\,{\cal B}} 
- \left( \,(\,\frac{4{\cal A}}{\omega _{0}} +
\frac{4{\cal C}^{2} + \omega _{0}^{2}}{4\omega _{0}\,{\cal B}}\,)^{2} 
- \frac{4{\cal A}}{{\cal B}} \,\right)^{1/2}  \,\right]
\,,
\end{eqnarray}
assuming $z \neq 0$.
The high frequency tail of this distribution decreases more rapidly
than that of the Boltzmann distribution, $e^{-\,\beta \omega }$.
On the other hand, at $\omega = 0$
\begin{equation}
f(0) = \frac{2}{\omega _{0}}\,\sqrt{\frac{{\cal A}}{{\cal B}}} \,.
\end{equation}
The average energy of the distribution is computed as
\begin{eqnarray}
\langle \omega  \rangle &\equiv &
\int_{0}^{\infty }\,d\omega \,\omega \,f(\omega )
\nonumber \\
&=&
\frac{\omega _{0}{\cal B}}{4{\cal A}}\,\left( \,
\frac{4{\cal A}}{\omega _{0}} + \frac{4{\cal C}^{2} + 
\omega _{0}^{2}}{4\omega _{0}\,{\cal B}}
\,\right) \,. \label{average energy} 
\end{eqnarray}
This average satisfies the relation,
\begin{equation}
\langle \omega  \rangle = \omega _{0}\,(\,\langle N \rangle + \frac{1}{2}\,)
\,,
\end{equation}
as expected.

The exceptional case of $z = 0$ with an exact exponential form
is actually important, as will be realized in thermal equilibrium. 
This is possible only when
\( \:
{\cal C} = 0 \,, \hspace{0.3cm} {\rm and} \hspace{0.3cm}
\sqrt{\,{\cal A}{\cal B}\,} = \omega _{0}/4 \,,
\: \)
giving
\begin{eqnarray}
f(\omega ) &=& \frac{e^{-\,\omega /T}}{T} 
\,, \hspace{0.5cm} 
{\rm with} \; T = 2{\cal B}
 \,, \\
\langle \omega  \rangle &=& T \,.
\end{eqnarray}

The normalized spectrum shape written in terms of a scaled energy,
$x = \omega /\langle \omega  \rangle$, is characterized by a single parameter
$\delta $;
\begin{eqnarray}
f_{0}(x\,; \delta ) &=& \frac{1}{\delta }\,e^{-\,x/\delta ^{2}}\,
I_{0}(\frac{\sqrt{1 - \delta ^{2}}}{\delta ^{2}}\,x) \\
&=&
\frac{1}{\pi \delta }\,e^{-\,x/\delta ^{2}}\,
\int_{0}^{\pi }\,d\theta \,\exp [\,-\,x\,\frac{\sqrt{1 - \delta ^{2}}}
{\delta ^{2}}\,\cos \theta \,] \,, 
\label{universal spectrum} 
\end{eqnarray}
with which the energy distribution is given by
\begin{eqnarray}
f(\omega ) &=& \frac{1}{\langle \omega  \rangle}\,
f_{0}\left(\,\frac{\omega }{\langle \omega  \rangle} \,; \delta\,\right ) 
\,, \\
\delta &=& 
2\sqrt{\,\frac{{\cal A}}{{\cal B}}\,}\,
\left(\,\frac{4{\cal A}}{\omega _{0}} + \frac{4{\cal C}^{2} + 
\omega _{0}^{2}}{4\omega _{0}\,{\cal B}} \,\right)^{-\,1} \,.
\end{eqnarray}
This function has the following limiting behavior:
\begin{eqnarray}
f_{0}(x\,; \delta ) &\rightarrow  & e^{-\,x} \; {\rm as} \;
\delta \:\rightarrow\: 1 \,, \\
f_{0}(x\,; \delta ) &\rightarrow & \frac{1}{\sqrt{2\pi \,x}}\,e^{-\,x/2}
\; {\rm as} \; \delta \:\rightarrow  \: 0 \,.
\end{eqnarray}
The first moment $\langle x \rangle$ happens to be the same in the
two limits of $\delta = 1 $ and $\delta = 0$.
The main difference of these two limiting distributions is in the low
$x$ part.
In fig.2 this universal function is plotted for a few choices of $\delta $.
We call $\delta $ the spectral shape parameter.

An important measure to explore the behavior of the quantum state
under the environment action is the effective entropy of the system
under study. We define as usual the entropy of the system as
(henceforth we omit the index $R$ for the reduced density matrix)
\begin{equation}
S = -\,{\rm tr}\;\rho \ln \rho \,, 
\end{equation}
where the trace operation is performed on the system variable $q$.
A useful device to compute the logarithmic matrix is to convert it to
a power series according to the formula like
\begin{equation}
S = -\,{\rm tr}\;
\left( \,\frac{d}{ds}\,\frac{1}{\Gamma (3-s)}\,\rho ^{3}\,
\int_{0}^{\infty }\,du\,u^{2-s}\,e^{-\rho u}\,\right)_{s = 1} \,.
\end{equation}
The arbitrary power 
\( \:
{\rm tr}\;\rho ^{n}
\: \)
is calculable,
\begin{eqnarray}
{\rm tr}\;\rho ^{n} &=& \int\,\prod_{i}^{n}\,dq_{i}\,
\rho ^{(R)}(q_{1}+q_{2}\,, q_{1}-q_{2})\cdots \rho ^{(R)}(q_{n}+q_{1}\,, 
q_{n}-q_{1}) \nonumber \\
&=&
\frac{(4{\cal A})^{n/2}}{(\,\sqrt{{\cal B}} + \sqrt{{\cal A}}\,)^{n} 
- (\,\sqrt{{\cal B}} - \sqrt{{\cal A}}\,)^{n}} \,.
\end{eqnarray}

A more efficient way to compute the entropy is first to observe,
with the presence of the power series expansion, independence
of the factor ${\cal C}$ in the entropy formula. This justifies neglect of
the ${\cal C}$ term (by setting ${\cal C} = 0$)
in the density matrix (\ref{density matrix}) for
computation of the entropy, and only for this purpose.
The next step is to identify this density matrix with ${\cal C} = 0$
as an equivalent harmonic oscillator system of frequency $\tilde{\omega }$
under a thermal bath of temperature $1/\tilde{\beta }$;
\begin{equation}
\sqrt{\,\frac{\tilde{\omega}}{\pi \,\coth 
(\tilde{\beta}\tilde{\omega} /2)}\,}\,
\exp [\,- \,\frac{\tilde{\omega}}{4}\,
\coth (\frac{\tilde{\beta}\tilde{\omega}}{2})\,
\xi _{i}^{2} - \frac{\tilde{\omega}}{4}\,
\tanh  (\frac{\tilde{\beta}\tilde{\omega}}{2})
\,X_{i}^{2}\,]  \,.
\end{equation}
The equivalence is possible only for 
\( \:
{\cal B} > {\cal A}
\: \)
and is established by the parameter relation,
\begin{eqnarray}
\tilde{\omega } = 4\,\sqrt{{\cal A}{\cal B}} \,, \hspace{0.5cm} 
\tanh \frac{\tilde{\beta }\tilde{\omega }}{2} = 
\sqrt{\frac{{\cal A}}{{\cal B}}} \,.
\end{eqnarray}
With this identification the equivalent temperature and the entropy are
given by
\begin{eqnarray}
\tilde{T} &=& \frac{4\,\sqrt{{\cal A}{\cal B}}}{\ln \left( \,
(\sqrt{{\cal B}} + \sqrt{{\cal A}})/
(\sqrt{{\cal B}} - \sqrt{{\cal A}}) \,\right)} \,, \\
S &=& \frac{\tilde{\beta }\tilde{\omega }}{e^{\tilde{\beta }\tilde{\omega }}
- 1} - \ln (\,1 - e^{-\,\tilde{\beta }\tilde{\omega }}\,) \nonumber \\
&=& \frac{1}{2}\, (x - 1)\ln \frac{x + 1}{x - 1} + \ln \frac{x + 1}{2}
\,, 
\label{entropy formula} 
\end{eqnarray}
with
\begin{equation}
x = \sqrt{\frac{{\cal B}}{{\cal A}}} 
\end{equation}
Clearly the effective frequency $\tilde{\omega }$ differs from
the average $\langle \omega  \rangle$ computed from the distribution
$f(\omega )$, eq.(\ref{average energy}).

When 
\( \:
{\cal A} \ll {\cal B} 
\: \) 
holds, considerable simplification follows;
\begin{eqnarray}
\tilde{\omega } &=& 4\sqrt{{\cal A}{\cal B}} \,, \\
\tilde{T} 
&\sim &
2\,{\cal B} \,, \\
S 
&\sim &
\frac{1}{2}\, \ln \frac{{\cal B}}{4{\cal A}} + 1 \,.
\end{eqnarray}

Comparison with the energy distribution function $f(\omega )$ makes it
clear that the thermal equilibrium state is possible only in the limit of
\begin{equation}
{\cal C} = 0 \,, \hspace{0.5cm} 
\frac{{\cal A}}{{\cal B}} \:\rightarrow  \: 0 \,, 
\end{equation}
in which $T \rightarrow 2{\cal B}$.
The second condition implies that
\begin{equation}
\frac{\omega _{0}}{T} = 2\,\sqrt{\frac{{\cal A}}{{\cal B}}} 
\:\rightarrow  \: 0 \,:
\end{equation}
The energy quantum $\hbar \omega _{0}$
must be much smaller than the thermal energy $k_{B}T$.

When ${\cal B} = {\cal A}$, 
the density matrix is factorizable into a product of 
\( \:
\psi (q)\,\psi ^{*}(q')
\: \),
since ${\rm tr}\;\rho ^{n} = 1$.
This is a pure quantum state, having the entropy
\( \:
S = 0 \,.
\: \)

Existence of the equivalent harmonic oscillator basis for 
${\cal B} \geq {\cal A}$ actually implies much more than a convenient
means of the entropy computation.
First, if ${\cal B} < {\cal A}$,
\( \:
{\rm tr}\;\rho ^{n} > 1
\: \)
for all $n\geq 2$, which means that despite an apparent probability
conservation ${\rm tr}\; \rho = 1$, some eigenvalues of $\rho $ are
negative; violation of the positivity of the density matrix.
Secondly, we have found the basis of the states that diagonalize
the density matrix at any instant of time: 
It is the Fock basis $|\,n \,\rangle $ of
the equivalent harmonic oscillator of frequency $\tilde{\omega }$
with diagonal elements;
\begin{equation}
\rho _{n} = 2\sinh (\frac{\tilde{\beta }\tilde{\omega }}{2})\,
e^{-\,\tilde{\beta }\tilde{\omega }\,(n + \frac{1}{2})} =
\frac{\sqrt{{\cal B} }- \sqrt{{\cal A}}}{2\sqrt{{\cal A}}}\,
\left( \,\frac{\sqrt{{\cal B} }- \sqrt{{\cal A}}}
{\sqrt{{\cal B} }+ \sqrt{{\cal A}}}\,\right)^{n}\,.
\end{equation}

It is thus important to check in any physical applications the inequality,
\begin{equation}
{\cal B} \geq {\cal A} \geq 0 \,,
\end{equation}
which is necessary for the positivity of the density matrix $\rho $.
In all practical applications
we shall discuss subsequently, the required inequality is obeyed, 
but when the picture of the local dissipative kernel is doomed to fail, 
this inequality, ${\cal B} \geq {\cal A}$ is also violated.
These occur either in the very short time behavior at
\( \:
\tau \leq 1/\Omega \,,
\: \)
or in the case of the low frequency cutoff, $\Omega \leq \omega _{0} $.
This has been checked in numerical computation of the harmonic
oscillator model mentioned in the next subsection.
Thus the model of the local friction fails very badly violating
the positivity condition,  when it does fail.
One has to be very cautious about the validity of the local approximation.

Finally, let us mention for comparison results that comes out when interaction
with the environment is switched off:
\( \:
r(\omega ) = 0 \; {\rm or} \; 
\eta = 0
\: \). 
The form of the density matrix (\ref{density matrix}) 
is the same as before with
\begin{eqnarray}
{\cal B} &=& \coth ^{2}(\frac{\beta _{0}\omega _{0}}{2})\,{\cal A} 
= \frac{1}{4\omega _{0}}\,
\coth (\frac{\beta _{0}\omega _{0}}{2})\,\frac{(\frac{\dot{u}(0)}{u(t)})^{2}}
{1 + (\,\frac{\dot{v}(0)}{\omega _{0}v(0)}\,)^{2}}\,, \\
{\cal C} &=&
\frac{1}{2\omega _{0}^{2}}\,(\frac{\dot{u}(0)}{u(t)})^{2}\,
\frac{\dot{v}(0)}{v(0)}\,
\frac{1}{1 + (\,\frac{\dot{v}(0)}{\omega _{0}v(0)}\,)^{2}} +
\frac{1}{2}\, \frac{\dot{u}(t)}{u(t)} \,.
\end{eqnarray}
When the system is in a pure quantum state initially with
 $\beta _{0} = \infty $,
it always remains in a pure state subsequently with
\( \:
{\cal A} = {\cal B} \,.
\: \)
Thus the entropy $S$ vanishes, while particle production can occur;
\( \:
\langle N \rangle > 0 \,.
\: \)

\vspace{0.5cm} 
\hspace*{-0.5cm}
{\bf \lromn 4 C. \hspace{0.3cm}
Time independent harmonic oscillator and
early time behavior
}

\vspace{0.5cm} 
A special case belonging to  our class of dynamical systems
is the independent collection
of simple harmonic oscillators in which $\omega _{R}^{2}$'s are
time independent. Although the simplest, this type of models is 
a fundamental model of the system-environment interaction and
used in various physical
situations: to name just two of them, quantum Brownian motion 
\cite{senitzky} --- \cite{lindenberg-west}. and
decoherence problem \cite{unruh-zurek}.
We shall recapitulate main computational results in our way of
approach, to the extent that comparison with application in the
next section is useful.
In the end of this subsection we discuss,
using these results for time independent $\omega _{R}^{2}$,
the early time behavior 
for the general case of time dependent $\omega _{R}^{2}(\tau )$.

The basic classical solutions in the case of a constant frequency are, with
\( \:
\omega _{R}^{2} = \omega _{0}^{2} \,, \\
\tilde{\omega }_{0}^{2} = \omega _{0}^{2} - \frac{\eta ^{2}}{4} \,, 
\: \)
\begin{equation}
u(\tau ) = \sin (\tilde{\omega} _{0}\,\tau ) \,, \hspace{0.5cm} 
v(\tau ) = \frac{\sin \left( \,\tilde{\omega} _{0}\,(t - \tau )\,\right)}
{ \sin (\tilde{\omega} _{0}\,t)} \,.
\label{classical for const freq} 
\end{equation}
Most of the relevant quantities can analytically be computed, except
the $\omega $ integral involving the factor $\coth \frac{\beta \omega }
{2}$ at finite $T$. 
The late time and/or the high temperature limits are fairly
simple, and to avoid too much complexity of formulas we only
record the high temperature and the late time behavior.
In these limits we may use
\begin{eqnarray}
&&
{\cal A} \:\rightarrow  \: \frac{1}{8B}\,(\frac{\dot{u}(0)}{u(t)})^{2}\,
\,, 
\\ &&
{\cal B} \:\rightarrow  \: \frac{A}{2} - \frac{C^{2}}{2B} \,, 
\\ && 
{\cal C} \:\rightarrow  \: 
\frac{C}{2B}\, \frac{\dot{u}(0)}{u(t)} -
\frac{1}{2}\, \left( \,\frac{\eta }{2} - \frac{\dot{u}(t)}{u(t)}\,\right)
\,.
\end{eqnarray}

What we mean here by high temperature actually covers a wide range of
interesting region. What should be compared with the temperature $T$
here is the energy scale of the system, $\omega _{0}$. The other quantity
$\eta $ is taken to be $O[g^{2}\,T]$ with $g$ some coupling $<1$.
Thus as far as $T \gg \omega _{0}$ is satisfied, one can regard the
system in the high temperature region.

A straightforward computation gives in the $T \rightarrow \infty $ limit;
\begin{eqnarray}
&&
{\cal A} \:\rightarrow  \: \frac{\omega _{0}^{2}}{8T} \,, 
\hspace{0.5cm} 
{\cal B} \:\rightarrow  \: \frac{T}{2} \,, 
\hspace{0.5cm} 
{\cal C} \:\rightarrow  \: 0 \,.
\end{eqnarray}
With these, physical quantities behave as
\begin{eqnarray}
\langle N \rangle &\rightarrow & \frac{T}{\omega _{0}} \,, 
\hspace{0.5cm} 
S \:\rightarrow  \: \ln \frac{T}{\omega _{0}} + 1 \,, 
\label{asymptotic in ho} 
\\
\langle \omega  \rangle &\rightarrow & 
T \,, \hspace{0.5cm} 
f(\omega ) \:\rightarrow  \: \frac{e^{-\omega /T}}{T} \,.
\end{eqnarray}
The entropy obeys the law,
\( \:
e^{S} \sim \langle N \rangle \,.
\: \)
Furthermore the density matrix given by ${\cal A} \,, {\cal B}$ and
${\cal C}$ here agrees with the thermal density matrix of the
environment temperature $T$ in the $T \rightarrow \infty $ limit.

These results indicate that the final state of the $q-$system is
insensitive to its initial state, as seen by independence of $\beta _{0}$,
and it is the universal state at the same temperature as the
environment. The approach of $\langle \omega  \rangle $ and 
$S$ to this asymptotia is fast, with
the exponential $e^{-\,\eta \,t }$ factor in correction terms,
as illustrated in fig.3 and fig.4 for a few choices of the friction $\eta $
and the initial system temperature $T_{0}$.
In these and all other subsequent numerical computations we set the
cutoff scale $\Omega = 50$. All other frequency and energy scales
are determined relative to this unit.
The final physical quantities are also insensitive to how
one treats the high frequency cutoff and even to the values of
$\eta $ and $T_{0}$.
The asymptotic values in these figures do agree with the analytic formula, 
eq.(\ref{asymptotic in ho}).
These are presumably an expected result, 
but computations here show the power and the simplicity of our method.

We finally discuss the behavior of physical quantities at early times, 
going back to the
general case of time dependent frequency $\omega _{R}^{2}(t)$.
Strictly speaking, our formalism only allows a precise estimate
of the behavior of the $q-$system at late times.
Nevertheless it would be of great interest to extrapolate results
to earlier times.
This is indeed the way people discuss the early time behavior in the
literature \cite{unruh-zurek}.
As noted previously, the very early time behavior at $\tau \leq 1/\Omega $
cannot be trusted in the approximation of the localized friction.
This means that to us the early time behavior refers to
\begin{equation}
\frac{1}{\Omega } < \tau \ll \frac{1}{\omega _{0}} \,.
\end{equation}

To this discussion we do not need to solve the classical equation for
$u(\tau )$ for the case of the time dependent frequency. 
Only this function for constant $\omega _{R}^{2}$, 
eq.(\ref{classical for const freq}) is needed, 
since time dependence appears in higher powers of $t$ for small $t$.
Thus the leading terms as $t \rightarrow 0^{+}$ are
\begin{eqnarray}
&&
A \,, \; B \,, \; C = O[t^{2}] \,, \\
&&
\frac{\dot{u}(0)}{u(t)} \rightarrow \frac{1}{t} \,, \hspace{0.5cm} 
\frac{\dot{v}(0)}{v(0)} \rightarrow -\,\frac{1}{t} \,.
\end{eqnarray}
Using these, we find that
\begin{eqnarray}
{\cal A} \: \rightarrow  \: \frac{\omega _{0}}{4}\,
\tanh (\frac{\beta _{0}\omega _{0}}{2})
\,, \hspace{0.5cm} 
{\cal B} \: \rightarrow  \: \frac{\omega _{0}}{4}\,
\coth  (\frac{\beta _{0}\omega _{0}}{2})
\,, \hspace{0.5cm} 
{\cal C} \:\rightarrow  \: -\,\frac{\eta }{2} \,.
\end{eqnarray}
Physical quantities behave as
\begin{eqnarray}
\langle \omega  \rangle &=& \omega _{0}\,(\,\langle N \rangle + \frac{1}{2}\,)
\: \rightarrow \:
\frac{\omega _{0}}{2}\,(\,1 + \frac{\eta ^{2}}{2\omega _{0}^{2}}\,)\,
\coth (\frac{\beta _{0}\omega _{0}}{2})
\,, \\
S &\rightarrow&  \frac{\beta _{0}\omega _{0}}
{e^{\beta _{0}\omega _{0}} - 1} - 
\ln \left( 1 - e^{-\,\beta _{0}\omega _{0}}\right) \,, \\
f(\omega ) &\rightarrow&
\frac{2}{\omega _{0}}\,\tanh (\frac{\beta _{0}\omega _{0}}{2})\,
\exp [\,-\,\frac{2\omega }{\omega _{0}}\,
(\,1 + \frac{\eta ^{2}}{2\omega _{0}^{2}}\,)\,
\tanh (\frac{\beta _{0}\omega _{0}}{2})\,]\,
\nonumber \\
&& 
\cdot I_{0}\left( \,\frac{\eta \omega }{\omega _{0}^{2}}\,
\sqrt{\,1 + \frac{\eta ^{2}}{4\omega _{0}^{2}}\,}\,
\tanh (\frac{\beta _{0}\omega _{0}}{2})\,\right)
\,. 
\end{eqnarray}
Without the system-environment interaction ($\eta = 0$) the quantum system
exhibits both vacuum and thermal fluctuation, while the entropy of
the system is given by the thermal contribution alone.

Thus the system starts at its given temperature $T_{0} = 1/\beta _{0}$.
Correction to this leading term arises only in powers of time $t$,
except the $\eta $ dependent terms in all quantities excluding
the entropy.
These $\eta $ dependent terms appear to show a
energy flow from the environment without accompanying entropy.
It persists even at $T_{0} = 0$, the case of a pure quantum state.

In the case of time independent harmonic oscillators the system finally
approaches the same thermal state of the environment smoothly.
We show in fig.3 and fig.4 result of numerical computations 
for constant $\omega _{R}^{2}$, to indicate how the occupation number
(fig.3) and the entropy (fig.4) evolve with time.

\vspace{1cm}
\section{Quantum system under periodic perturbation}

\vspace{0.5cm} 
\hspace*{0.5cm} 
We finally specialize to the case in which a quantum system field
$\varphi $ is coupled to a coherent field oscillation
\( \:
\xi (t) \propto \cos (m_{\xi }t)
\: \). 
(For a lack of notation we use the same, already used Greek letter
$\xi $ to denote the external oscillator field in this section.)
The coherent field may be viewed as an aggregate of zero-momentum
particles with some kind of precise coherence.
$m_{\xi }$ is thus the mass of these bosons.
In our present work this field oscillation is regarded as given, 
hence we do not discuss field damping due
to particle production of the $\varphi-$field.
Our main concern here is in the effect of thermal bath on the 
$\varphi-$particle production caused by $\xi-$oscillation.

Under this circumstance the harmonic oscillator variable is
\( \:
q_{k} = 
\frac{1}{\sqrt{\,2\omega _{k}\,V\,}\,}\,(\,a_{k} + a^{\dag }_{-k}\,) \,, 
\: \)
where the creation and the annihilation operators  are those of
Fourier $\vec{k}-$ mode ($\propto e^{i\vec{k}\cdot \vec{x}}$) of
the system field $\varphi (x)$.
For the oscillator-system interaction we take the quartic coupling
given by
\begin{equation}
\frac{1}{2}\, g^{2}\xi ^{2}\varphi ^{2} \,,
\end{equation}
primarily because this case has been analyzed in considerable detail
\cite{fkyy95} 
without taking into account the environment effect.
The $\vec{k}-$mode variable $q_{k}$ then obeys the evolution equation,
\begin{equation}
\left( \,\frac{d^{2}}{d\tau ^{2}} + \omega ^{2}(\tau )\,\right)
\,q_{k}(\tau ) = 0 \,, \hspace{0.5cm} 
\omega ^{2}(\tau ) = \vec{k}^{2} + m_{\varphi }^{2}
+ g^{2}\xi _{0}^{2}\,\cos^{2}(m_{\xi }\tau )
\,. \label{original mahieu} 
\end{equation}
$\xi _{0}$ is the amplitude of oscillation.
The standard form of this type of equation is called the Mathieu
equation and is usually written in dimensionless units;
\begin{eqnarray}
&&
\left( \,\frac{d^{2}}{dz^{2}} + h + 2\theta \,\cos (2z)\,\right)
\,q(z) = 0 \,, \\
&&
\hspace*{0.5cm} 
z = m_{\xi }\tau \,, \hspace{0.5cm} 
h = \frac{\vec{k}^{2}+m_{\varphi }^{2}}{m_{\xi }^{2}} + 
2\theta \,, \hspace{0.5cm} 
\theta = \frac{g^{2}\xi _{0}^{2}}{4m_{\xi }^{2}} \,.
\end{eqnarray}

It is well known that this quantum system exhibits instability in infinitely
many band regions \cite{landau-lifschitz m} of the parameters of
\( \:
(\,|\vec{k}| \,, g\xi_{0}\,) \,,
\: \) 
or $(h\,, \theta )$
.
It has been realized that this instability gives rise to particle production
\cite{baryogenesis-reheating-rev}.
The most important recent development is that for large amplitude
oscillation particle production and associated field decay is
greatly expedited, typically ending in a time scale of 
O[10 --- 100] times $1/m_{\xi }$
\cite{linde et al 94},\cite{holman 95},\cite{fkyy95}.
However, in all initial investigations so far no systematic estimate of the
environmental effect has been attempted.
This is precisely what we wish to do in the present section.

Taking over the formalism developed in the previous sections,
we ask how the classical solution $u(\tau )$ and $v(\tau )$ evolve
with time $\tau $,
from which we can compute all physical quantities of the quantum 
$\varphi-$system.
Let us first recall that these functions satisfy the Mathieu 
equation modified by the frequency shift and the friction:
\begin{equation}
\left( \,\frac{d^{2}}{d\tau ^{2}} + \omega ^{2}(\tau ) +
\delta \omega ^{2} - \frac{\eta ^{2}}{4}\,\right)\,u(\tau ) = 0 \,.
\label{modified mathieu} 
\end{equation}
The frequency shift is in proportion to the cutoff:
\( \:
\delta \omega ^{2} \propto \Omega \,, 
\: \)
hence is divergent in field theory models. As usual in renormalizable
field theories it is more appropriate to regard the renormalized mass
\( \:
m_{\varphi }^{2} + \delta \omega ^{2}
\: \)
as a physical parameter.
This function $u(\tau )$ is related to the original mode function 
$q(\tau )$ by
\begin{equation}
u(\tau ) = e^{\eta \tau/2 }\,q(\tau ) \,.
\end{equation}
Limiting our analysis to modes within the instability band of the modified
Mathieu equation, we note that
two linearly independent solutions are either growing or decaying
according to
\begin{equation}
e^{\lambda m_{\xi }\tau }\,P(\tau ) \,, \hspace{0.5cm} 
e^{-\,\lambda m_{\xi }\tau }\,R(\tau ) \,, 
\end{equation}
where $\lambda > 0$ is the dimensionless growth rate and both $P(t)$ and
$R(t)$ are periodic with period of $ 2\pi /m_{\xi }$.
Needed solutions must satisfy the boundary condition,
\( \:
u(0) = 0 \,, \; v(t) = 0 \,, 
\: \)
hence these are linear combinations of the growing and the decaying
solutions. With the normalization given in eq.(\ref{normalization}), 
leading asymptotic behaviors of these are
\begin{equation}
u(t) \sim e^{\lambda m_{\xi }t}\,\tilde{P}(t) \,, \hspace{0.5cm} 
\dot{v}(t) = - \,\frac{\dot{u}(0)v(0)}{u(t)} \sim 
e^{-\,\lambda m_{\xi }t}\,\tilde{R}(t) \,,
\end{equation}
with both $\tilde{P}(t)$ and $\tilde{R}(t)$ bounded functions.

Note that the growth rate $\lambda $ here is related to the solution
of the modified Mathieu equation, eq.(\ref{modified mathieu}), and
not of the original one (\ref{original mahieu}).
Thus we should keep in mind that the growth rate $\lambda $ does
depend on the friction $\eta $:
\( \:
\lambda = \lambda (\eta ) \,.
\: \)
Suppose for instance that the renormalized $\varphi-$mass vanishes:
\( \:
m_{\varphi }^{2} + \delta \omega ^{2} = 0 \,.
\: \)
The $\eta $ term in eq.(\ref{modified mathieu}) then has an effect of
lowering the band level, since the $h$ parameter in the standard Mathieu
equation is given by
\begin{equation}
\tilde{h} = h - \frac{\eta ^{2}}{4m_{\xi }^{2}}
= \frac{\vec{k}^{2}}{m_{\xi }^{2}} + 2\theta 
- \frac{\eta ^{2}}{4m_{\xi }^{2}}
\end{equation}
in this case.
Although it is in general difficult to estimate the influence of 
the friction $\eta $
on the growth rate $\lambda (\eta )$, the general trend is obvious:
The lowering effect of the band level tends to increase the growth rate.
In the extreme case of a very large $\eta $ the $h$ parameter may become
negative, and the parameter is clearly in the instability region.
Needless to say, in perturbation theory we later work with, the leading
$\lambda $ term is independent of $\eta $.
Of course, the real criterion of the parametric amplification
is derived only after we work out time evolution of physical quantities.

The exponential growth of the original mode function $q(\tau )$
is possible for\\
 $2\lambda(\eta )\, m_{\xi } > \eta $.
We observe two competitive factors for the growth:
the rate of the parametric amplification $2\lambda(\eta )\, m_{\xi }$ 
against the friction $\eta $.
As will be shown in subsequent computations of physical quantities, 
the friction, when it is small, does
act to diminish the parametric particle production, but does not
wipe out the parametric effect. Thus, instead of the blocking factor,
it is more appropriate to view  the friction as
the inverse time scale for the system to be driven towards thermalization.
Hence if the friction is small enough, the parametric amplification
never loses against the friction.

Although implicit, it is important to remember $v(\tau )$ depends on
$t$ through the boundary condition.
Thus it is more appropriate to write this as $v(\tau \,; t)$. 
An example of these classical solutions, $u(\tau )$ and $v(\tau )$ 
normalized at $\tau = t$ and $\tau = 0$ respectively, are shown in fig.5.

In subsequent discussion we present both analytic and numerical computation
of time evolution of physical quantities. 
In analytical computation we can give only the limiting behavior.
On the other hand, detailed numerical computation has been performed,
which covers  a variety of situations.
Some of these results are shown in figures.
In analytic estimate we concentrate on the asymptotic late time behavior 
for which
\( \:
\tau \gg {\rm max.} \;(\,1/\eta \,, 1/(\lambda m_{\xi })\,)
\: \).
With the asymptotic behavior of the classical solution, $u(\tau ) $ and $ 
v(\tau )$, the three quantities that appear in the reduced density matrix 
(\ref{density matrix}) of the $\varphi-$system are
\begin{eqnarray}
{\cal A} &=& \frac{1}{8K}\,(\frac{\dot{u}(0)}{u(t)})^{2}
\,, 
\\ 
{\cal B} &\rightarrow&  \frac{A}{2} - \frac{C^{2}}{2K} \,, 
\\ 
{\cal C} &\rightarrow  &
\frac{C}{2K}\, \frac{\dot{u}(0)}{u(t)}
- \frac{1}{2}\, \left( \,\frac{\eta }{2} - \frac{\dot{u}(t)}{u(t)}\,\right)
\,.
\end{eqnarray}
The quantities $K\,, A \,, C$ are already defined in eq.
(\ref{k-equation} --- \ref{last c-eq}) and
have asymptotic behaviors,
\begin{equation}
A = O[1] \,, \hspace{0.5cm} C \leq  
O[\,{\rm Max.\;} (\,e^{-\,\lambda m_{\xi }t}\;,e^{-\,\eta t/2})\,] 
\,, \hspace{0.5cm} 
K = O[\,{\rm Max.\;} (\,e^{-\,2\lambda m_{\xi }t}\;,e^{-\,\eta t})\,] \,.
\end{equation}
Thus, we conclude that ${\cal A} \,, {\cal B} \,, 
{\cal C}$ behave asymptotically as
\begin{equation}
{\cal A} = O[\,{\rm Min.}\;(\,1\,, e^{-\,2\lambda m_{\xi }t + \eta t}\,)\,] 
\,, \hspace{0.5cm} {\cal B} = O[1] \,, \hspace{0.5cm} 
{\cal C} = O[1] \,.
\end{equation}
We omitted dimensional scales of these quantities such as
$\omega _{0}$ and $\eta $, since these differ from case to case,
depending on the parameter range of
\( \:
T \,, \; \omega _{0} \,, \; m_{\xi } \,, \; \eta \,.
\: \)
For the reference frequency we take
\( \:
\omega _{0}^{2} = \omega _{R}^{2}(0) \,:
\: \)
the initial value of the frequency.
We also note that these asymptotic forms are sensitive to the initial
$\varphi-$state, dependence of $T_{0} = 1/\beta _{0}$, only via the $K$ term
of eq.(\ref{k-equation}).

The formulas so far are valid for any response weight $r(\omega )$.
We now specialize to the environment system that may be approximated
by the localized friction with
\( \:
r(\omega ) = \frac{\eta }{\pi }\,\omega \; \times 
\: \)
(high $\omega-$cutoff).
In particular, we are very much interested in the 
high and low temperature limits of various quantities.
As in the preceding case of constant frequency the high temperature
here effectively means that 
\( \:
T \gg {\rm Max.}\;(\,\omega _{0} \,, m_{\xi } \,, \eta \,, \Omega \,)
\,.
\: \)

First, in the high temperature limit of $T \:\rightarrow  \: \infty $
the frequency integral is well convergent and one has
\begin{eqnarray}
&&
\int_{0}^{\infty }\,d\omega \,\coth \frac{\omega \beta }{2}\,
\omega \,\cos (\omega (\tau - s)) \sim \frac{2\pi }{\beta }\,
\delta (\tau - s) \,, 
\\
&& \hspace*{-1.5cm}
\left( \,A \,, B \,, C\,\right) \sim 
\frac{2\eta }{\beta }\,\int_{0}^{t}\,d\tau \,e^{-\,\eta (t - \tau )}\,
\left( \,(\frac{u(\tau )}{u(t)})^{2} \,, (\frac{v(\tau )}{v(0)})^{2}
\,, \frac{u(\tau )v(\tau )}{u(t)v(0)}\,\right) \,.
\end{eqnarray}
Thus in terms of powers of $T$,
\( \:
{\cal A} = O[1/T] \,, \hspace{0.3cm}
{\cal B} = O[T] \,, \hspace{0.3cm}
{\cal C} = O[1] \,.
\: \)
Let us first focus on the case of strong parametric resonance for which
\( \:
2\lambda m_{\xi } \gg \eta 
\: \)
such that the effect of parametric resonance is well operative.
In this case
\begin{eqnarray}
\frac{{\cal B}}{\eta T} &\sim& 
\int_{0}^{t}\,d\tau\,e^{-\eta (t - \tau )}\,
(\frac{u(\tau )}{u(t)})^{2} 
\nonumber \\
&-& \left( \,\int_{0}^{t}\,d\tau \,
e^{-\,\eta (t - \tau )}\,(\frac{u(\tau )v(\tau )}{u(t)v(0)})\,\right) ^{2}
\left( \,\int_{0}^{t}\,d\tau\,e^{-\eta (t - \tau )}\,
(\frac{v(\tau )}{v(0)})^{2}\,\right)^{-1} \,, 
\\
{\cal A} &\sim& \frac{1}{8\eta T}\,(\frac{\dot{u}(0)}{u(t)})^{2}\,
\left( \,\int_{0}^{t}\,d\tau\,e^{-\eta (t - \tau )}\,
(\frac{v(\tau )}{v(0)})^{2}\,\right)^{-1} \,, 
\\
{\cal C} &\sim& 
-\,\frac{1}{2}\, \left( \,\frac{\eta }{2} - \frac{\dot{u}(t)}{u(t)}\,\right) 
\,.
\end{eqnarray}
Note that the expression for ${\cal C}$ is  different from the
constant $\omega _{R}^{2}$ case discussed at the end of the previous section,
because in the present case $u(t) $ exponentially grows.
The integrals here are crudely estimated by noting that
$u(\tau )$ increases rapidly at the time scale of $1/(\lambda m_{\xi })$,
much shorter than $1/\eta $, starting from $\tau = 0$, while $v(\tau )$
decreases at the same time scale.
We checked this behavior numerically by integrating the Mathieu equation.
Thus
\begin{eqnarray}
{\cal A} &\approx & 
\frac{1}{8}\,\frac{\lambda m_{\xi }}{\eta T}\,e^{\eta t}\,
(\frac{\dot{u}(0)}{u(t)})^{2} \,, 
\\
{\cal B} &\approx &
\frac{\eta T}{2\lambda m_{\xi }} \,.
\end{eqnarray}

From these asymptotic forms we find at high temperatures
\begin{eqnarray}
\langle \omega  \rangle &\sim  & \omega _{0}\, \langle N \rangle 
\; \approx \;
\frac{\eta \omega _{0}^{2}\,T}{2\lambda m_{\xi }}\,
e^{-\,\eta t}\,(\frac{u(t)}{\dot{u}(0)})^{2}\,
\left( \,1 + 
\frac{1}{\omega _{0}^{2}}\,(\,\frac{\eta }{2} - \frac{\dot{u}(t)}{u(t)}\,)^{2}
\,\right) 
\,, \\
S &\approx &
\ln \,\left( \,\frac{\eta T}{\lambda m_{\xi }}\,|\frac{u(t)}{\dot{u}(0)}|
\,e^{-\eta t/2}\,\right)
\,, \\
f(\omega ) &\approx &
\frac{1}{\sqrt{\,2\pi \,\langle \omega  \rangle\,}}\,
\frac{e^{-\,\omega /(\,2\langle \omega  \rangle\,)}}{\sqrt{\omega }} \,.
\label{asymptotic dist}
\end{eqnarray}
Both the number $\langle N \rangle$ and the average energy
$\langle \omega  \rangle$ increase exponentially with time, the rate
being somewhat diminished from the value taken without environment.
Note also that a rather non-trivial enhancement factor $\:\propto  \: T$
appears in the prefactor.

We show in fig.6 --- fig.8 time evolution of various quantities resulting
from numerical computation at
relatively high temperatures in the cases of both the
small and the large friction.
The adopted parameters relevant to the original Mathieu equation are
\( \:
h = 1 \,, \hspace{0.3cm} 2\theta = 0.75 \,,
\: \)
and correspondingly
\( \:
k  \sim 0.5\,m_{\xi } \,, \hspace{0.3cm}
g\xi _{0} \sim 1.2\,m_{\xi } \,.
\: \)
This belongs to the intermediate $\xi-$amplitude region giving
$\lambda \sim 0.185$ in vacuum.
The global growth rate computed from the modified Mathieu equation is
\( \:
2\lambda (\eta )\,m_{\xi } - \eta \sim 0.17 
\: \)
in the case of $\eta = 0.2$,
which fits well with numerical results of time evolving physical
quantities.
Observe in these figures that the case of a large friction given here by
$\eta = 1.0$ does not yield the parametric amplification,
as independently checked by
\( \:
2\lambda (\eta )\,m_{\xi } - \eta \sim -\,0.69 < 0
\: \).
An outline of how we numerically computed 
\( \:
{\cal A} \,, {\cal B} \,, {\cal C}
\: \)
is described in Appendix.
An example of time evolution of these basic quantities is shown
in fig.9.

The energy distribution (\ref{asymptotic dist}) 
exaggerates the low energy part, because
it does not yield the correct $\omega \rightarrow 0$ limit.
A better approximation is
\begin{eqnarray}
f(\omega ) &=& 
\frac{1}{\sqrt{\,2\pi \,\langle \omega  \rangle\,}}\,
\frac{e^{-\,\omega /(\,2\langle \omega  \rangle\,)}}
{\sqrt{\omega + \omega _{c}}} \,, \\
\omega _{c} &\sim & \frac{2}{\pi }\,\frac{\omega _{0}^{2}{\cal B}}
{4\,{\cal C}^{2} + \omega _{0}^{2}} \approx 
\frac{1}{\pi }\,\frac{\eta T}{\lambda m_{\xi }}\,
\left( \,1 + 
\frac{1}{\omega _{0}^{2}}\,(\,\frac{\eta }{2} - \frac{\dot{u}(t)}{u(t)}\,)^{2}
\,\right)^{-\,1}
 \,.
\end{eqnarray}
The low energy cutoff $\omega _{c} \ll \langle \omega  \rangle$
(the average).
Although the energy distribution differs from the thermal one, 
the dominant component is in the high energy side.

In this model the total system combined with the environment 
may gain energy only from the oscillating field $\xi (t)$.
The energy input rate from the coherent oscillator is estimated by
\begin{equation}
\frac{d}{dt}\,\langle E_{{\rm tot}} \rangle = -\,
\frac{g^{2}\xi _{0}^{2}}{16{\cal A}}\,m_{\xi }\,\sin (2m_{\xi }t)
\,.
\end{equation}
The rate itself contains oscillating factors, but as shown in fig.10,
the integrated energy 
\begin{equation}
\langle E_{{\rm tot}} \rangle =
-\,\int_{0}^{t}\,d\tau \,
\frac{g^{2}\xi _{0}^{2}}{16{\cal A}(\tau )}\,m_{\xi }\,\sin (2m_{\xi }\tau )
\,,
\end{equation}
always increases after transient times of $t \leq O[\,\pi /(2m_{\xi })\,]$.
A deep reason how this comes about is the cooperative interference
of the two oscillating components,
\( \:
\sin (2m_{\xi }t)
\: \)
and the one in
\( \:
1/{\cal A} \propto (\,\frac{u(t)}{\dot{u}(0)}\,)^{2} \,.
\: \)
If $u(t)$ monotonically increases without oscillation unlike in this case
of the parametric resonance, there is no
net gain of the energy $\langle E_{{\rm tot}} \rangle$ from the oscillator.
With regard to this it is interesting to note that in the case of a large
friction the quantity $1/{\cal A}$ does not exponentially grow, but
has a regulated oscillatory behavior, which results in a linear, instead
of the exponential, growth
of the energy flow $\langle E_{{\rm tot}} \rangle$ as illustrated in
the inlet of fig.10.

In a sense the quantum system of the parametric amplifier maintains
in one way or another its quantum nature even at high temperatures,
provided that 
\( \:
2\lambda (\eta )\,m_{\xi } - \eta > 0 \,.
\: \)
This is in sharp contrast to the case of simple harmonic oscillator put
in thermal oscillator bath, for which we have seen that the late time
behavior is the same thermal state of the environment.
It might be possible to experimentally observe a quantum behavior
of the parametric amplifier at high temperatures.

Despite the suppression
of the growth rate from $2\lambda m_{\xi }$ to $2\lambda m_{\xi } - \eta $,
the environment interaction does not erase the parametric effect if
\begin{equation}
2\lambda(\eta )\, m_{\xi } > \eta \,.
\label{parametric condition} 
\end{equation}
The derivation of this growth condition
constitutes one of the most important conclusions of the present work.
Let us look into the meaning of this condition
that guarantees the exponential rate of particle production
in thermal bath.
Since $\eta $ is the relaxation rate of $\varphi-$field disturbance
towards thermalization,
the condition simply implies that the thermal relaxation 
never catches up the parametric amplification for a sufficiently
small friction.
At the same time the condition implies
that there exists a critical strength of the friction $\eta $
above which the parametric effect does not occur.
The equation 
\( \:
2\lambda (\eta )\,m_{\xi } = \eta 
\: \)
thus gives the critical strength of the friction constant $\eta _{c}$.
We wish to stress that although the suppression of the growth rate might
have be anticipated from a naive consideration \cite{kasuya-kawasaki},
it was not clear in the past \cite{dolgov-kirilova comm} 
how one can verify the parametric amplification
itself, starting from quantum mechanical principles.
In our view the significance of the friction $\eta $ is in its
role as the relaxation rather than destruction of the coherence.
Hence if the relaxation time scale is larger than the amplification time
scale, the parametric effect wins over the decoherence due to the environment.

The growth condition eq.(\ref{parametric condition}) can be quantified.
We shall base our estimate of this inequality
on the field theory model that incorporates the quartic Yukawa 
\( \:
\frac{1}{2}\, g^{2}\xi ^{2}\varphi ^{2}
\: \)
of the oscillator coupling and the system-environment coupling of the form
\( \:
\frac{1}{2}\, g_{Y}m_{\xi }\, \varphi \chi ^{2} \,.
\: \)
(We parametrized the previous $\mu $ by a dimensionless 
$g_{Y} = \mu /m_{\xi }$). Furthermore for simplicity we assume that
the renormalized $\varphi $ mass vanishes.
Since we take the lowest non-trivial order in perturbation theory,
the $\eta $ dependence of $\lambda $ may be ignored in the discussion here.
The inequality condition (\ref{parametric condition})
is then, using (\ref{friction in field th})
\begin{equation}
\frac{k}{n} > \frac{g_{Y}^{2}}{64\lambda }\,m_{\xi } \,, \hspace{0.5cm} 
\frac{1}{n} = 
\exp \left( \,\frac{\sqrt{\,k^{2} + 4m^{2}\,}}{2T}\,\right) - 1 \,.
\end{equation}
Physical processes that contribute to the dissipative friction here
are
\( \:
\chi + \varphi \leftrightarrow \chi 
\: \)
that becomes dominant at high temperatures.
If instead the dominant process of dissipation is 
\( \:
\varphi \rightarrow \chi + \chi \,, 
\: \)
then the following constraint may be modified substantially.
We hope to cover this case in subsequent investigation that involves
non-linear dissipation.
The growth rate $\lambda $ does depend on the mode variable, 
specifically the
$(k \,, g\xi _{0})$ value within the instability band.
It was however shown in ref \cite{fkyy95} that for the large amplitude of
\( \:
g\xi _{0} \gg m_{\xi } \,,
\: \)
the average 
\( \:
\bar{\lambda } \approx 0.2 \,.
\: \)
The inequality  thus limits the momentum value of produced particles.
Since the quantity $k/n$ in the left hand side at a fixed $k$ decreases
as the temperature $T$ increases, the momentum range becomes smaller
as $T$ increases. But for $k \geq T$, $k/n$ rapidly increases, as
the momentum $k$ increases.
Thus this constraint is not very stringent for $k \geq T$.

Thus $\eta = 2\lambda m_{\xi }$ gives the critical friction below which
the parametric amplification occurs, as illustrated in fig.6 --- 8.
Although the case of the overdamping $ \eta > 2\lambda m_{\xi }$ is
not our main concern, we shall write down results in this case,
which have some interesting features worth to mention.
Numerical results in this case are illustrated with the largest
$\eta $ value ($\eta = 1.0$) in fig.6 -- fig.8.
At high temperatures the basic physical quantities behave
asymptotically;
\begin{eqnarray}
\langle \omega  \rangle &\sim & \omega _{0}\,\langle N \rangle \approx 
\frac{T}{2}\,
\left[ \,1 + \frac{4\omega _{0}^{2}}{\eta ^{2}}\,
\left( \,1 + \frac{1}{\omega _{0}^{2}}
\,(\frac{\dot{u}(t)}{u(t)}\,)^{2}\,\right)\,\right] 
\,, \\
S &\approx & \ln \frac{2T}{\eta } + 1 \,,\\
f(\omega ) &\approx &
\frac{1}{\langle \omega  \rangle}\,
f_{0}\left( \,\frac{\omega }{\langle \omega  \rangle}
\; ; \; \frac{2\omega _{0}\,T}{\eta \langle \omega  \rangle}\,\right)
\,.
\end{eqnarray}
As expected, there is no effect of the parametric amplification.
The $\delta $ parameter that characterizes the spectral shape is
\begin{equation}
\delta \:\rightarrow  \: \frac{4\omega _{0}}{\eta }\,
\left[ \,1 + \frac{4\omega _{0}^{2}}{\eta ^{2}}\,
\left( \,1 + \frac{1}{\omega _{0}^{2}}
\,(\frac{\dot{u}(t)}{u(t)}\,)^{2}\,\right)\,\right]^{-1} \,.
\end{equation}
Time evolution of this spectral shape parameter is depicted in fig.8 by
$\eta = 1.0$ line for the large friction. The oscillatory behavior around
$\delta  = 0.9$ is clearly observed.
Thus even a large friction cannot destroy the quantum nature of the 
parametric amplifier: In a sense the system never completely
equilibrates.

We next turn to the low temperature limit of $T \:\rightarrow  \: 0$
in which $\coth \frac{\beta \omega }{2} \rightarrow 1$ in the 
$\omega-$integral.
In the reheating problem this case is more relevant, since the
universe supercools after inflation.
In this discussion we need the high frequency cutoff at $\Omega $ and
first utilize with $\epsilon \equiv 1/ \Omega \,, $
\begin{eqnarray}
\int_{0}^{\infty }\,d\omega \,\omega \,\cos \omega (\tau - s)\,
e^{-\,\epsilon  \omega} =
\frac{d}{d(s - \tau )}\,\frac{s - \tau }{(s - \tau )^{2} + \epsilon ^{2}}
\,.
\end{eqnarray}
Singular terms may arise as $\epsilon \rightarrow 0^{+}\,. $ 
We must retain the leading logarithmic singularity such as
\( \:
\ln (\frac{1}{m_{\xi }\,\epsilon }) \sim 
\ln (\frac{\Omega }{m_{\xi }})
\: \),
because this reflects accumulated effect to the high frequency cutoff
$\Omega $.
We shall explain estimate of the basic quantities only in the simpler case of
the small friction, $\eta < 2\lambda m_{\xi }$, and state 
the final result for $\eta > 2\lambda m_{\xi }$.
Numerically both the small and the large friction cases have been
computed.

Let us illustrate concretely some integration:
\begin{equation}
\hspace*{-1cm}
A = 
\frac{\eta }{\pi }\,\int_{0}^{t}\,d\tau \,\int_{0}^{t}\,ds\,
\frac{u(\tau )u(s)}{u(t)^{2}}\,
\exp [\,-\,\frac{\eta }{2}\,(2t - \tau - s)\,]\,
\frac{d}{d(s - \tau )}\,\frac{s - \tau }{(s - \tau )^{2} + \epsilon ^{2}}
\,.
\end{equation}
We partial integrate with respect to $s$ and expand around $s = \tau $
to estimate this type of integral. The leading term of $A$ above
is the surface term in the partial integration, which gives 
\begin{eqnarray}
A \approx  \frac{\eta }{\pi }\,\int_{0}^{t}\,d\tau \,\frac{u(\tau )}{u(t)}\,
e^{-\,\eta (t - \tau  )/2}\,\frac{t - \tau  }
{(t - \tau )^{2} + \epsilon ^{2}} \sim \frac{\eta }{\pi }\,
\ln (\frac{\Omega }{m_{\xi }}) \,.
\end{eqnarray}
The rest of the integral in $A$ does not contain the logarithmic term 
in the $\epsilon \rightarrow 0$ limit, and may be ignored.

Similarly 
\begin{eqnarray}
\hspace*{-2cm}
B &=& 
\frac{\eta }{\pi }\,\int_{0}^{t}\,d\tau \,\int_{0}^{t}\,ds\,
\frac{v(\tau )v(s)}{v(0)^{2}}\,\exp [\,-\,\frac{\eta }{2}\,
(2t - \tau - s)\,]\,\frac{d}{d(s - \tau )}\,
\frac{s - \tau }{(s - \tau )^{2} + \epsilon ^{2}} \nonumber \\
&\approx &
\frac{\eta }{\pi }\,\int_{0}^{t}\,d\tau \,\frac{v(\tau )}{v(0)}\,
e^{-\,\eta (\,2t - \tau \,)/2}\,\frac{\tau }{\tau ^{2} + \epsilon ^{2}}
\;\approx \;
\frac{\eta }{\pi }\,e^{-\,\eta t}\,
\ln (\frac{\Omega }{m_{\xi }}) \,, \\
C &\approx  &
\frac{\eta }{m_{\xi }t}\,e^{-\eta t/2} \,.
\end{eqnarray}
From these
\begin{eqnarray}
K &\sim & e^{-\,\eta t}\,E \,, \\
E &=& \left[\,\frac{\eta }{\pi }\ln (\frac{\Omega }{m_{\xi }}) + 
\frac{\omega _{0}}{2}\coth (\frac{\beta _{0}\omega _{0}}{2})\,
\left( \,1 + \frac{1}{\omega _{0}^{2}}(\,\frac{\eta }{2} +
\frac{\dot{v}(0)}{v(0)}\,)^{2}\,\right)\,\right]
\,, \\
{\cal A} &\sim & 
\frac{1}{8E}\,
e^{\eta t}\,(\frac{\dot{u}(0)}{u(t)})^{2}
\,, \\
{\cal B} &\sim & 
\frac{\eta }{2\pi }\,\ln (\frac{\Omega }{ m_{\xi }}) \,, \\
{\cal C} &\sim & 
-\,\frac{1}{2}\, \left( \,\frac{\eta }{2} - \frac{\dot{u}(t)}{u(t)}\,\right) 
\,.
\end{eqnarray}
The reason why the dependence on the initial state factor $\beta _{0}$
disappears is dominance of the 
\( \:
\ln (\frac{\Omega }{m_{\xi }})
\: \) term in $K$.

Physical quantities such as the average number of produced particles and
the entropy are asymptotically 
\begin{eqnarray}
\langle \omega  \rangle
&\sim & \omega _{0}\,\langle N \rangle \;\sim \;
\frac{E}{2}\,
\left( \,1 + \frac{1}{\omega _{0}^{2}}\,
(\,\frac{\eta }{2} - \frac{\dot{u}(t)}{u(t)}\,)^{2}\,\right)\,e^{-\,\eta t}
\,(\frac{u(t)}{\dot{u}(0)})^{2}
\,, \\
e^{S} &\sim & 
\sqrt{\,\frac{\eta E}{\pi }\,\ln (\frac{\Omega }{m_{\xi }})\,}\,
e^{-\,\eta t/2}\,\,|\,\frac{u(t)}{\dot{u}(0)}\,| \,.
\end{eqnarray}
This result markedly differs from what we know without the environment
effect:
A pure quantum state has zero entropy, while interaction with environment
gives a randomness in agreement with
\( \:
S \approx \frac{1}{2}\, \ln \langle N \rangle \,,
\: \)
for large $\langle N \rangle$.
The energy distribution is given by
\begin{eqnarray}
f(\omega ) &\approx &
\frac{1}{\sqrt{\,2\pi \,\langle \omega  \rangle\,}}\,
\frac{e^{-\,\omega /(\,2\langle \omega  \rangle\,)}}
{\sqrt{\omega + \omega _{c}}} \,, \\
\omega _{c} &\approx & 
\frac{\eta }{\pi^{2} }\,\ln (\frac{\Omega }{m_{\xi }})\,
\left( \,1 + \frac{1}{\omega _{0}^{2}}\,
(\,\frac{\eta }{2} - \frac{\dot{u}(t)}{u(t)}\,)^{2}\,\right)^{-\,1}
\,.
\end{eqnarray}

In the large friction of $\eta > 2\lambda m_{\xi }$ at low temperatures
a similar analysis gives
\begin{eqnarray}
\langle \omega  \rangle &\approx & \frac{\eta }{2\pi }\,\ln (\frac{2\Omega }
{\eta }) \,, \\
S &= & S(x) \,, \hspace{0.5cm} 
x \approx \frac{1}{\pi }\,\sqrt{\,8\ln (\frac{2\Omega }{\eta })\,}
\,, 
\end{eqnarray}
with $S(x)$ defined by eq.(\ref{entropy formula}).

The formulas written here are valid for 
\( \:
\eta \,\ln (\frac{\Omega }{ m_{\xi }}) \gg \omega _{0} \,.
\: \)
When this does not hold, the factor, 
\( \:
\eta \,\ln (\frac{\Omega }{ m_{\xi }}) \,, 
\: \)
should be replaced by some intrinsic frequency scale of the $\varphi-$system.
Our results at low temperatures indicate that the difference from the
high temperature case is described by a replacement,
\( \:
\frac{T}{\lambda m_{\xi }} \:\rightarrow  \: 
\ln (\frac{\Omega }{m_{\xi }}) \,, 
\: \)
ignoring subleading terms.
Thus the temperature plays a minor role in the parametric amplification.
The crucial condition is the inequality,
\( \:
2\lambda(\eta )\, m_{\xi } > \eta \,.
\: \)

Results of numerical computation at $T = 0$
are shown in fig.11 --- fig.13.
Global features of time evolution at low temperatures do not differ much 
from the high temperature case, as anticipated by analytic results above.
Thus if time is displaced, most physical quantities at different
temperatures look the same.
This is a surprising, and an important result of our work, indicating that 
the parametric effect is insensitive to the environment temperature
except through the rate
\( \:
2\lambda (\eta )\,m_{\xi } - \eta \,.
\: \)
In fine details this universal feature is somewhat spoiled since the
amount of the time displacement is not the same for the average energy
$\langle \omega  \rangle$ and the entropy $S$.

As advocated in another model \cite{kls-2}, our results indicate that
the energy spectrum at the instant of particle creation
deviates from the thermal one: it is neither of the form of Bose condensation.
However,
the spectral shape does not necessarily persist till the last moment, 
since interaction among created particles neglected in the present work
can establish thermalization, as discussed in ref \cite{fkyy95}.

Our computation in the present work has been limited to a single mode
by specifying $k $ and $\xi _{0}$ (the amplitude of $\xi $ oscillation).
To assess more fully effect of the parametric amplification, one
must further mode-sum over $\vec{k}$ in all possible instability
bands. A consequence of this mode-sum might be to erase the oscillatory
behavior in time evolution of the average energy and the entropy.
Thus although the quantum behavior caused by the parametric amplifier 
persists in medium, the mode-sum may or may not erase the bizarre oscillatory
behavior observed in a single mode calculation here.
We have also performed a sample computation when the parameters of
the coherent oscillation are outside the instability band and inside
a stability region.
The exponential growth is not observed in this case as expected, but
the system does not amalgamate to the environment, either.
The quantum behavior such as a persistent 
oscillatory behavior is observed, thus
it seems that the periodic perturbation keeps the quantum system
from losing its quantum nature even in thermal bath.

Although it is not our main topic to deal with the reheating problem after
inflation, it might be of some interest to comment on this.
Our approach is based on the premise that separation between
the system and the environment is evident.
In the preheating stage after inflation, especially prior to copious
production of $\varphi $ particles, the separation is obvious.
Indeed, the relevant environment is the vacuum at zero temperature.
Whether the Bose condensation may or may not occur can be decided
at this stage.
On the other hand, after explosive particle production the environment
changes. Clearly the new environment consists of created particles due to
the inflaton decay prior to thermalization.
We cannot discuss this stage in our framework.
This is regrettable since this stage is the most important to
the later cosmological evolution.
We also note that we have neither discussed how the inflaton decay
is completed: here the back reaction caused by particle production
is very important.
Some recent works on particle interaction after their creation are found
in \cite{khlebnikov 96-1}, \cite{son 96}, \cite{allashverdi et al}. 

Although our achievement on the reheating problem is very modest,
our result on the environment effect to the parametric amplification may
be of considerable use in other cosmological problems.
These are the Polonyi or the moduli problem.
These  weakly interacting fields are thought to decay after long periods
of matter dominance, thus potentially giving a serious entropy crisis 
at their decay \cite{polonyi}.
Their oscillation lasts long enough in cosmic medium, thus it is important
to estimate the environment effect.

\vspace{1cm}
\section{Summary}

\vspace{0.5cm} 
\hspace*{0.5cm} 
Since a great many have been described in the present paper, it would be
useful to summarize here what we think we achieved in this work.
Our work addresses how to quantify the effect of the environment on
the behavior of a quantum system and can be classified into three parts:
The first concerns
general formalism to characterize and approximate the system-environment
interaction; the second concerns the effect of the environment on the
general dynamical system of time dependent harmonic oscillators;
the third concerns the effect of the environment on dynamics of harmonic
oscillator with periodic frequency.

Our results on these three issues are as follows.\\
On general formalism of the system-environment interaction:
\begin{itemize}
\item 
We extended the influence functional method by introducing the spectral
weight $r(\omega )$, called the response weight. 
For the bilinear coupling between the system and the environment
variables the correlation function that appears in the influence functional
is identified to the Green's function in the environment medium,
for instance the real-time thermal Green's function for the environment
in thermal equilibrium. The response weight completely characterizes
the system-environment interaction.
\item 
We developed two methods of modeling the system-environment interaction.
The first one is a meromorphic approximation to the response weight,
somewhat akin to the resonance approximation in the low energy scattering
problem.
The second one is a perturbation theory in terms of the interaction strength.
The Ohmic behavior at low frequencies, $r(\omega ) \propto \omega $, 
is a common feature of models discussed in the present work.
\item 
We introduced a class of non-Gaussian field theory models of the interaction
Lagrangian density of the form ${\cal L}_{{\rm int}} = -\,
\frac{1}{2}\, \mu \varphi \chi ^{2}$
between the system field $\varphi $ and the environment field $\chi $.
Finite temperature computation in this field theory model has
been worked out at one loop level.
How to model this case as an extended Ohmic model was briefly
discussed.
\end{itemize}

\hspace*{-0.5cm}
On the time dependent harmonic oscillator in thermal bath:
\begin{itemize}
\item 
The late time behavior of the system under the environment effect
is described by an approximation to the correlation function;
the dissipative part of the correlation given by a local form,
\( \:
\alpha _{I}(\tau ) \sim \eta \,\delta '(\tau ) \,,
\: \)
where the friction constant $\eta $ describes the inverse time
scale of relaxation of the system variable towards thermalization.
Correspondingly the response weight
\( \:
r(\omega ) = \frac{\eta }{\pi }\,\omega \,f(\frac{\omega }{\Omega })
\,, \: \)
with $f(x)$ decreasing rapidly as $x \rightarrow \infty $.
$\Omega $ is the cutoff of the high frequency part.
We identified the response weight and the friction constant in
the two models of the system-environment interaction.
\item 
We solved in a closed form 
the problem of how time dependent harmonic oscillators
behave asymptotically at late times: The reduced density matrix of the system
under the effect of thermal environment is determined
in terms of two quantities,
a single classical solution $p(\tau )$ obeying
\( \:
\stackrel{..}{p} - \eta \,\dot{p}+ \omega ^{2}(\tau )p = 0
\: \)
with 
\( \:
p(0) = 0 \,, \; \dot{p}(0) = 1 \,,
\: \)
and the response weight.
\item 
Physical quantities such as the energy distribution,
the entropy of the system, and the total population number in oscillator
levels have been explicitly computed.
The resulting energy spectrum is given by a universal function of the
scaled energy, $x =$ energy / (average energy), having the form,
\( \:
\exp [-\,x/\delta ^{2}]\,I_{0}(\,x\,\sqrt{1 - \delta ^{2}}/\delta ^{2}\,)
/\delta \,, 
\: \)
where the time dependent parameter $1 - \delta $ 
(with $0\leq  \delta \leq 1$)
describes a degree of departure from thermal equilibrium of the system.
\end{itemize}

\hspace*{-0.5cm}
On the environment effect against the parametric oscillator:
\begin{itemize}
\item 
For the first time it was verified that the parametric amplification
can occur in medium, using the influence functional method.
The average energy and the total number of population
exponentially increases with time, if the growth rate of the parametric
amplification in medium is larger than the friction $\eta $,
interpreted to be a measure of the relaxation rate towards thermalization
of the system variable.
The critical friction for the exponential growth has also been given, 
above which the parametric effect does not occur.
\item 
The resulting energy distribution deviates from the thermal Boltzmann
distribution, having a long tail of the high energy component 
characterized by an exponentially increasing average energy and
an exponentially decreasing $\delta $ parameter.
\item 
Late time behavior of the average energy, the entropy, and the
energy spectrum is insensitive to the environment temperature
if time is displaced. This is true even including the $T = 0 $ vacuum.
\end{itemize}

After submission of this paper for publication, we became aware
of some related works on the reheating and the preheating problems
after inflation \cite{boyanovsky et al 96-7}, \cite{kls 96},
\cite{khlebnikov 96-2}.

\newpage
\begin{center}
{\bf Acknowledgment}
\end{center}

This work has been supported in part by the Grand-in-Aid for Science
Research from the Ministry of Education, Science and Culture of Japan,
No. 08640341 and 08740185.
One of us (M.H.) acknowledges the hospitality  of the SLAC theory group 
while part of this work has been done during his stay there.

\newpage
\appendix
\section{How to numerically compute 
${\cal A} \,, {\cal B} \,, {\cal C}$}

\vspace{0.5cm} 
The basic quantities we must know before computing physical quantities 
are 
${\cal A} \,, {\cal B} \,, {\cal C}$
introduced in Section \lromn 4.
These contain some factors like $u(t)$ in denominators, and
although zeros of these functions cancel in the end, it would be
much better to avoid these zeros in intermediate steps of calculation.
We achieve this by introducing, and expressing 
\( \:
{\cal A} \,, {\cal B} \,, {\cal C}
\: \)
in terms of, supplementary functions that
obey simple differential equations with definite boundary conditions.
We briefly describe here the method to do this. 

Let us first recall that the fundamental function $u(\tau )$ obeys 
the following relations;
\begin{eqnarray} 
  \ddot{u}(\tau ) &=& f(\tau )\,u(\tau ) \,, \\
  u(0) &=& 0 \,, \hspace{0.5cm} 
  \dot{u}(0) = 1 \,.
\end{eqnarray}
For notational simplicity we replaced the time dependent frequency 
$\omega^{2} (\tau )$ by $-\,f(\tau )$.
The secondary function $v(\tau \,; t)$ is then
\begin{eqnarray}
  v(\tau \, ; t) &\equiv& u(\tau)\int_t^\tau \frac{d\tau'}{u^2(\tau')} \,,\\
  v(t \, ; t) &=& 0 \,, \hspace{0.5cm} 
  v(0 \, ; t) = -1 \,.
\end{eqnarray}
We distinguish the two arguments, $\tau $ and $ t$, to indicate that
$\tau $ is a time variable and $t$ is the end time at which we
set the boundary condition. The dot always means the $\tau $ derivative.
The normalization here was chosen such that the Wronskian 
\begin{equation}
u(\tau )\,\dot{v}(\tau \,; t) - \dot{u}(\tau )\,v(\tau \,; t) = 1 \,.
\end{equation}

We further define two more functions;
\begin{eqnarray}
  h(\tau \, ; t) &\equiv& u(t)\,(\, - \,\dot{v}(\tau) + \frac{\eta}{2}\,)
   \nonumber \\
  &=& - \,u(t)\left(\, \dot{u}(\tau )\int_t^\tau
  \frac{d\tau'}{u^2(\tau')} + \frac{1}{u(\tau)}
  - \frac{\eta}{2} \,\right) \,.
\end{eqnarray}
Note that $t \geq \tau $.
These functions are shown to satisfy
\begin{eqnarray}
  \partial_t^2 h(\tau \,; t) &=& f(t) \,h(\tau \, ; t) \,, \\
  h(0 \,;0) &=& -1 \,, \hspace{0.5cm} 
  \partial_t h(0 \,;0) = \frac{\eta}{2}\,.
\end{eqnarray}
Finally we define 
\begin{eqnarray}
  c(t) &\equiv& -\,u(t)\,\int_0^t\,d\tau\, 
  v(\tau \,; t)\,\cos(\omega \tau)\,e^{\eta\tau/2 } \,, \\
  s(t) &\equiv& -\,u(t)\int_0^t\,d\tau\,
  v(\tau \,;t)\,\sin(\omega \tau)\,e^{\eta\tau/2 }\,, \\
  g(t) &\equiv& \lim_{\tau \rightarrow 0}\, h(\tau \,; t) \,,
\end{eqnarray}
which are shown to obey
\begin{eqnarray}
  \partial_t^2 g(t) &=& f(t)\,g(t) \,, \\
  g(0) &=& -\,1 \,, \hspace{0.5cm} 
  \partial_t g(0) = \frac{\eta }{2} \,, \\
  \partial_t^2 c(t) &=& f(t)\,c(t) + \cos(\omega t)\,
  e^{\eta\tau/2 } \,, \\
  c(0) &=& 0 \,, \hspace{0.5cm} 
  \partial_t c(0) = 0 \,, \\
  \partial_t^2 s(t) &=& f(t)\,s(t) + \sin(\omega t)\,
  e^{\eta\tau/2 }\,, \\
  s(0) &=& 0 \,, \hspace{0.5cm} 
  \partial_t s(0) = 0 \,.
\end{eqnarray}

We introduce the following multiplication and the integral transform
to simplify subsequent expressions;
\begin{eqnarray}
  {\cal D}_0 &\equiv& 
  \frac{\omega_0}{2} \coth(\frac{\beta_0 \omega_0}{2})\,\times \,,  \\
  {\cal D}_\omega\,(\,F\,) &\equiv& 
  \int_0^\infty\, d\omega\, r(\omega)\,\coth(\frac{\beta \omega}{2})\,
  \times F(\omega ) \,.
\end{eqnarray}

Basic functions are then 
\begin{eqnarray}
\hspace*{-1cm}  
{\cal A} &=& \frac{e^{\eta t}}
  {8 \left(\, {\cal D}_0 (u^{2}(t) + \frac{g^{2}(t)}{\omega_0^2})
     + {\cal D}_\omega\, (\,c^{2}(t) + s^{2}(t)\,)\, \right)} \,, \\
\hspace*{-1cm}
{\cal B} &=& 4{\cal A}\, e^{-\,2\eta t}\,
     \left[\,
    \frac{{\cal D}_0^2}{\omega_0^2} 
    + {\cal D}_0\,{\cal D}_\omega \,
    \left(\, (\,
      \dot{u}^{2}(t) + \frac{\dot{g}^{2}(t)}{\omega_0^2}\,)\,
      (\,c^{2}(t) + s^{2}(t)\,) \,
    \right. \right. \nonumber \\
\hspace*{-1cm}    &+& \left. 
    (\,u^{2}(t) + \frac{g^{2}(t)}{\omega_0^2}\,)
    (\,\dot{c}^{2}(t) + \dot{s}^{2}(t)\,) 
    - 2(\,\dot{u}(t)u(t)
      + \frac{\dot{g}(t)g(t)}{\omega_0^2}\,)\,
      (\,\dot{c}(t)c(t) + \dot{s}(t)s(t)\,)\,
    \right) \nonumber \\
    &+& \left. \{\,{\cal D}_\omega\,(\,\dot{c}^{2}(t) +
      \dot{s}^{2}(t)\,) {\cal D}_\omega\,(\,c^{2}(t) + s^{2}(t)\,) 
      - (\,{\cal D}_\omega\,(\,\dot{c}(t)c(t) + \dot{s}(t)s(t))\,)^2\}\,
      \right]
\,, \\
\hspace*{-1cm}
{\cal C} &=& 4{\cal A}\, e^{-\, \eta t}\,
  \left(\,
    {\cal D}_0 \,(\,\dot{u}(t)u(t) +  \frac{\dot{g}(t)g(t)}{\omega_0^2})
     + {\cal D}_\omega\,(\,\dot{c}(t)c(t) + \dot{s}(t)s(t)\,) \,
   \right)  - \frac{\eta}{4} \,.
\end{eqnarray}
We thus need to solve four linear differential equations for
\( \:
u(\tau ) \,, \; g(t) \,, \;
c(t) \,, \; s(t) \,,
\: \)
that obey different boundary conditions, to get 
${\cal A} \,, {\cal B} \,, {\cal C}$. Some of these equations
contain inhomogeneous terms that depend on $\omega $.

\newpage

\vspace{0.2cm} 

\newpage
\begin{center}
{\bf Figure captions}
\end{center}

\vspace{0.5cm} 
{\bf Fig.1}

Response weight in the field theory model at finite temperature.
One loop result of 
\( \:
r(\omega ) = 2\,\Im \,\Pi (\omega )
\: \)
with $\Im \,\Pi (\omega )$ the imaginary part of the self-energy
diagram is shown for a parameter set;
\( \:
T = 50 \,, k = 5 \,,   \mu = 1 
\: \)
in the energy unit of $m$ (the mass of decaying particle).
Also shown by the dotted line is the approximation to this weight
given by eq.(3.\ref{approximate field th model}) in the text.

\vspace{0.5cm} 
{\bf Fig.2}

Energy distribution given by the universal form,
\( \:
f_{0}(x \,; \delta ) 
\: \)
of eq.(4.\ref{universal spectrum}),
plotted against a scaled energy of $x= \omega /\langle \omega  \rangle$ 
for three choices of
$\delta = 1.0$ (the dotted line), $ 0.5 $ (the solid line), $0.2$
(the broken line). 

\vspace{0.5cm} 
{\bf Fig.3}

Time evolution of 
populated number (= average energy $/\omega _{0} - \frac{1}{2}$) 
of harmonic oscillator levels
for various choices of the parameters, 
$\eta $ (the friction) and $T_{0}$ (the system temperature),
all at the environment temperature $T = 100$; \\
\( \:
(\eta \,, \; T_{0}) = (0.1 \,, 0) \,, \; 
(0.2 \,, 0 ) \,, \; 
(1.0 \,, 0) \,, \; 
(0.2 \,, 100) \,, \; 
(0.2\,, 150) 
\: \) \\
in the energy unit of $\omega _{0}$, given by the designated lines with
a, b, c, d, e, respectively.

\vspace{0.5cm} 
{\bf Fig.4}

Time evolution of entropy for harmonic oscillator for the same set of
parameters as in Fig.3.

\vspace{0.5cm} 
{\bf Fig.5}

Time evolution of the fundamental solutions for the modified Mathieu
equation, normalized at $\tau = t$ to give
\( \:
\frac{u(\tau )}{u(t)}
\: \)
(the solid line), 
and at $\tau = 0$ to give
\( \:
\frac{v(\tau )}{v(0)}
\: \)
(the dotted line).
Relevant parameters are
\( \:
h = 1 \,, \hspace{0.3cm} 2\,\theta = 0.75 \,, \hspace{0.3cm}
\eta = 0.2\,m_{\xi } \,, \hspace{0.3cm} 
t = 50/m_{\xi } \,,
\: \)
and the time scale is measured by $1/m_{\xi }$.

\vspace{0.5cm} 
{\bf Fig.6}

Time evolution (in the same time scale of Fig.5) of 
occupation number (= average energy $/\omega _{0} - \frac{1}{2}$) 
of oscillator levels of reference frequency $\omega _{0}$
in the case of the parametric oscillator with
\( \:
h = 1 \,, \; 2\theta = 0.75 \,.
\: \)
Three different situations are compared:
$\eta = 0$ or in vacuum by the broken line, 
the small friction case with
\( \:
(\eta\,, T) = (0.2 \,, 100) 
\: \)
by the solid line,
and the large friction case with
\( \:
(1.0 \,, 100)
\: \)
by the dotted line. The system temperature is all set at $T_{0} = 0$,
the ground state.

\vspace{0.5cm} 
{\bf Fig.7}

Time evolution (in the same time scale of Fig.5) 
of entropy in the case of the parametric
oscillator with
\( \:
h = 1 \,, \; 2\theta = 0.75 \,.
\: \)
Three different situations of the friction $\eta $ and the environment
temperature $T$ are compared:
$(\eta \,, T) = (0.05 \,, 100 )$ by the broken line, 
\( \:
(0.2 \,, 100)
\: \)
by the solid line,
and
\( \:
(1.0 \,, 100)
\: \)
by the dotted line. The system temperature is all set at $T_{0} = 0$,
the ground state.

\vspace{0.5cm} 
{\bf Fig.8}

Time evolution of the spectral shape parameter $\delta $.
Two cases of the small and the large friction at a high temperature
($T = 100$) for the parametric oscillator ($h  = 1.0$ and
$2\,\theta = 0.75$) 
are compared to the simple harmonic oscillator 
(with $\eta  = 0.2 )$ depicted by the dotted line;
the small friction case ($\eta = 0.2 $) given by
the broken line, and the large friction case 
$(\eta = 1.0 )$ given by the solid line.

\vspace{0.5cm} 
{\bf Fig.9}

Time evolution of the basic quantities 
\( \:
{\cal A} \,, \hspace{0.3cm} {\cal B} 
\: \)
in figure 9A and 
\( \:
{\cal C}
\: \)
in figure 9B,
for 
\( \:
h = 1 \,, \hspace{0.3cm} 2\,\theta = 0.75 \,, \hspace{0.3cm}
\eta = 0.2 \,, \hspace{0.3cm} T = 100 \,.
\: \)
The time scale is measured by the angular oscillation frequency of
the parametric oscillator, $1/m_{\xi }$.

\vspace{0.5cm} 
{\bf Fig.10}

Time evolution (in the same time scale of Fig.5)
of integrated energy flow from the parametric oscillator
into the total system including the environment.
Three cases are compared:
\( \:
\eta = 0 \; ({\rm isolated \; system})
\: \)
by the solid line,
$(\eta\,, T) = (0.2 \,, 100)$ by the broken line,
and $(1.0 \,, 100)$ by the dotted line,
all at the system temperature 
$T_{0} = 0$. Inside the inlet is shown the case of the large friction of
$(1.0 \,, 100)$ in a linear enlarged scale.

\vspace{0.5cm} 
{\bf Fig.11}

Time evolution (in the same time scale of Fig.5) of 
populated number of oscillator levels of a reference frequency $\omega _{0}$
in the case of the parametric oscillator with
\( \:
h = 1 \,, \; 2\theta = 0.75 \,.
\: \)
Three different situations in vacuum ($T_{0} = 0$)
are compared to the high temperature case by designated letters:
$(\eta \,, \; T \,, \; T_{0}) = \; $
(a) $\, (0.2 \,, \; 0 \,, \; 0) \;$;
(b) $\, (0.2 \,, \; 100 \,, \; 0) \;$;
(c) $\, (0.2 \,, \; 0 \,, \; 100) \;$;
(d) $\, (1.0 \,, \; 0 \,, \; 0) \;$;
(e) $\, (0 \,, \; 0 \,, \; 0)\,.$

\vspace{0.5cm} 
{\bf Fig.12}

Time evolution (in the same time scale of Fig.5) 
of entropy for the set of parameters;
$(\eta \,, \; T \,, \; T_{0}) = \; $
(a) $\, (0.2 \,, \; 0 \,, \; 0) \;$;
(b) $\, (0.2 \,, \; 100 \,, \; 0) \;$;
(c) $\, (0.2 \,, \; 0 \,, \; 100) \;$;
(d) $\, (1.0 \,, \; 0 \,, \; 0) \,.$

\vspace{0.5cm} 
{\bf Fig.13}

Time evolution (in the same time scale of Fig.5)
of the spectral shape parameter $\delta $ with the friction
$\eta = 0.2$, at high ($T = 100$ by the broken line) and at low
($T = 0$ by the solid line) temperatures, taken 
for the parametric oscillator of
\( \:
h = 1 \,, \; 2\theta = 0.75 \,.
\: \)


\begin{thebibliography}{99}
\bibitem{dolgov-kirilova comm}
Some skepticism against the parametric effect in medium was
raised in the first paper in reference [2].
\end{thebibliography}
\end{document}